\documentclass[fleqn,usenatbib]{mnras}

\usepackage{newtxtext,newtxmath}
\usepackage{tabularx}

\usepackage[T1]{fontenc}

\DeclareRobustCommand{\VAN}[3]{#2}
\let\VANthebibliography\thebibliography
\def\thebibliography{\DeclareRobustCommand{\VAN}[3]{##3}\VANthebibliography}

\usepackage{graphicx}	
\usepackage{amsmath}	
\usepackage[normalem]{ulem}
\usepackage{makecell}
\usepackage{soul}      
\usepackage{xcolor}  

\title[Exocomets in TESS Sectors 1-26]{A Search for Transiting Exocomets in TESS Sectors 1-26}

\author[A. Norazman et al.]
{\parbox{\textwidth}{\Large
{
Azib Norazman,$^{1,2}$\thanks{E-mail: azib.norazman@warwick.ac.uk}
Grant M. Kennedy,$^{1,2,3}$
Ann Marie Cody,$^{4}$
Daniel Giles,$^{4,5}$
Samuel Gill,$^{1,2}$\\
Ethan Kruse$^{5}$}\\[-7pt]}
\\
$^{1}$Department of Physics, University of Warwick, Gibbet Hill Road, CV4 7AL, UK\\
$^{2}$Centre for Exoplanets and Habitability, University of Warwick, Gibbet Hill Road, Coventry CV4 7AL, UK\\
$^{3}$Malaghan Institute of Medical Research, Gate 7, Victoria University, Kelburn Parade, Wellington, New Zealand\\
$^{4}$SETI Institute, 339 N Bernardo Ave Suite 200, Mountain View, CA 94043, USA\\
$^{5}$NASA Goddard Space Flight Center, 8800 Greenbelt Road, Greenbelt, MD 20771, USA
}

\date{Accepted XXX. Received YYY; in original form ZZZ}

\pubyear{2025}

\begin{document}
\label{firstpage}
\pagerange{\pageref{firstpage}--\pageref{lastpage}}
\maketitle

\begin{abstract} 

\noindent We present a search for single photometric exocomet transits using a magnitude-limited sample of stars observed by the TESS primary mission. These events are asymmetric, with a sharp ingress and more gradual egress expected because the comet tail trails behind the coma. Our goals are to estimate the occurrence rate of exocomet transits, and given sufficient numbers comment on whether the host stars are biased towards being A/F spectral types, as suggested by a previous survey with \textit{Kepler} data. We recovered the previously identified exocomet transit with TESS around $\beta$~Pic (TIC~270577175) and identified three additional main sequence systems with exocomet-like transits (TIC~280832588, TIC~73149665, and TIC~143152957). We also identified one exocomet candidate around a giant star (TIC~229790952) and one around a probable supergiant (TIC~110969638). We find a total occurrence rate of $2.64\times10^{-4}$ star$^{-1}$\,year$^{-1}$, much higher than \textit{Kepler}'s rate of $6.7 \times 10^{-6}$ star$^{-1}$\,year$^{-1}$. Some of this difference may be because our rate includes a correction for detection efficiency, where the \textit{Kepler} search did not. However, with only a handful of detections in each survey, the rates are also very uncertain. In contrast to the \textit{Kepler} search, we find two candidate hosts that may be G types, but the spectral types would be better supported with spectroscopic follow-up. Primarily, we conclude that exocomet-like transits are very rare at 0.1\%-1\% transit depth levels, and that higher precision photometry to detect and characterise shallower transits effectively is the most likely path to more detections and stronger statistical conclusions. 


\end{abstract}




\section{Introduction}

Comets are a stunning display in the night sky. They are a well-known constituent of our Solar system and are composed primarily of ice and rock, with sizes ranging from $\sim$ 1km to a few hundreds of kilometres. This class of planetesimals did not get accreted by nearby forming planets and reside in distinct reservoirs in many planetary systems \citep[e.g.][]{Wyatt2008,Krivov2010}, e.g. the Asteroid belt, the Edgeworth-Kuiper belt, and the Oort cloud in our Solar System. As a result of dynamical interaction with stars and planets, planetesimals can be perturbed towards our inner Solar System \citep{Morbidelli2005}, and as they approach the Sun, they start to sublimate, where material is lost from the nuclei, creating distinct tails that we can see from Earth. Planet formation theories suggest that comets and other planetesimals should be a common component of all stellar systems \citep{Johansen2014}. 

Exocomets are the extra-solar analogues of Solar System comets, with a broad definition of small sublimating bodies that orbit other stars. As individual exocomet bodies are detected by photometric/spectroscopic transits, there is little in the way of further classification \citep{kiefer2014}, compared to Solar system comets which are classified by composition and dynamics \citep{Strom2020, fitzsimmons_matra_isos_exocomets_2023}. Exocomets are thought to originate in Asteroid/Kuiper belt analogues of their host stars, commonly known as debris discs (sometimes referred to as exocometary belts or exo-Kuiper belts, though here we do not consider detections of the belts themselves to be detections of exocomets). Exocomets are currently the only class of minor bodies that we can observe individually in extrasolar planetary systems\footnote{though dust-emitting asteroids observed to transit white dwarfs have also been detected \citep[e.g.:][]{Vanderburg2015,Guidry2021,Bhattacharjee2025}}, where their detection is attributed to enlarged comae/tails resulting from the release of dust and gas as they pass the sublimation radius (i.e. we do not detect the nuclei directly). Historically, the main method for the detection of exocomets in extrasolar systems is transit spectroscopy. The first evidence to suggest exocometary activity was variations in the absorption features of the CaII H and K lines around the young, A-type star Beta~Pictoris (i.e.: $\beta$~Pic) \citep{ferlet1987, ferlet1993, kiefer2014}, which were interpreted as transiting exocomet comae. Similar varying spectral features were also detected around the star HD 172555 \citep{kiefer_2014_exocomets_hd172555}, and these two stars are the most convincing exocomet host systems. Other exocomet candidate systems are not as well characterised due to less data available to validate those detections. We refer the reader to Iglesias et al. (in review) and \cite{Strom2020} and the references therein for a list of exocomet detections.

The second method to detect exocometary bodies is through broadband transit photometry. Here, we observe the ``shape'' of an exocomet as a sharp ingress as the coma occults the host star, followed by a gradual increase beyond the mid-transit point back to normal stellar flux levels as the tail passes across the star \citep{Lecavalierdesetangs1999}. The capability for photometric observations of exocomets is driven by the rapid advancement of large-scale transiting planet surveys, where work beyond the original aims of these missions has led to exciting discoveries. \textit{Kepler} was the first mission with sufficient instrumental precision to detect exocomets. \cite{Rappaport2018} conducted a visual survey of all \textit{Kepler} light curves, finding two systems that strongly matched exocomet transit shape predictions in \cite{Lecavalierdesetangs1999} in the light curves of KIC 3542116 and KIC 11084727. In addition, \cite{Kennedy2019} built on these discoveries by developing an automated search method that detects single transits across the entire \textit{Kepler} dataset, characterising candidates by their asymmetry, with the premise that automated methods are an efficient alternative to churn through large volumes of data. With this method, they recovered the two transits in \cite{Rappaport2018} and discovered a third system showing asymmetric transit behaviour around KIC 8027456 (HD 182962), where the transit depth for these events were about 0.1\%. The two key findings of \cite{Kennedy2019} were that i) the exocomet candidates detected were around young A/F-type stars, and ii) the exocomet detection rate was approximately $7\times10^{-6}$ per star per year (but they did not attempt to quantify or correct for the efficiency of their search). Given that dynamical activity (e.g. scattering of planetesimals by planets) is thought to be greater for younger stars, detections around younger stars are perhaps consistent with theories of planetary system formation and evolution \citep{Wyatt2008}.



The most recent wave of space-based photometry with the Transiting Exoplanet Survey Satellite  \citep[TESS;][]{Ricker2015} has enabled the search for more exocomets. TESS is an all-sky survey that observes overlapping patches of the sky (i.e. ``sectors'') for 27 days each, monitoring hundreds of thousands of stars in each sector with its $24^{\circ} \times 96^{\circ}$ field of view. $\beta$~Pic was observed relatively early on in the TESS primary mission, where exocomets were detected in sectors 5 and 6 \citep{Zieba2019}, confirming it to be the first star with observations of exocomets in both spectroscopy and photometry. Observations of $\beta$~Pic in subsequent TESS sectors also yielded more detections, allowing an estimation of the exocomet size distribution for the system \citep{lecavalierdeseetangs2022}. The size distribution is similar to that of Jupiter family comets (where comets' orbits are primarily influenced by the presence of Jupiter), suggesting that collisional fragmentation is the likely mechanism that shapes the size distribution of exocomets around $\beta$~Pic. In addition, HD 172555 also showed a potential exocomet detection with the CHaracterising ExOPlanet Satellite (CHEOPS) \citep{kiefer_cheops_hd172555_2023}, where removal of the $\delta$-scuti pulsations revealed an event that matches the exocomet transit models of \cite{Lecavalierdesetangs1999}. These two stars are therefore the only ones that have detections in both photometry and spectroscopy.


Although spectroscopic and photometric detections of exocomets provide evidence for these objects around $\beta$~Pic and HD~172555, they are not necessarily detecting the same bodies: photometric detections sample a wider range of orbital distances (up to 1 au \citep{Zieba2019}), while spectroscopic observations detect events much closer to the host star (e.g.: $<$ 0.15 au for $\beta$~Pic \citet{kiefer2014,kennedy2018_exocomet_fitting,Beust1989-PaperIX-Theoretical-Results,Lagrange1989-Evidence-structure-of-infalling-gas}). In addition, no exocomet has yet been detected simultaneously through both methods, making it difficult to establish whether the same objects can be observed through both techniques. An example of this challenge was presented by photometric observations around the star 5 Vul, a star hypothesised as a good target star from its stellar properties and previously detected spectroscopic observations \citep{Montgomery2012,Iglesias2018,ExocometsRebollido}. Observations of 5 Vul with CHEOPS for 45 hours did not yield any exocomet detections \citep{rebollido_2023_cheops_5vul}, which may be related to the inability to detect photometric transits for objects that are close to the star, or that more distant objects do not sublimate or transit, or that there were simply no transits during that CHEOPS observation window.

The goal of this work is to build on the results in \cite{Kennedy2019} and conduct a large-scale survey for exocomets with TESS using data from its primary mission. Overall, exocomet detections appear to come from young A/F type stars \citep{Strom2020}. This trend has been seen in spectroscopic surveys, but the samples of these surveys have typically been biased towards debris disc hosts, and therefore biased towards A/F types. On the other hand, photometric surveys are not as biased, since missions such as \textit{Kepler} and TESS are focused on observing large samples of stars ($10^5-10^7$ stars) to look for planet transits, and therefore target selection is not inherently biased towards debris disc host systems. 

The all-sky nature of TESS's observing strategy is promising as it provides a magnitude-limited sample of stars, such that there are no explicit biases in the data. Here, we aim to search for exocomet-like transits in TESS data and quantify their occurrence, and given sufficient detections draw some conclusions about whether these are more or less likely to appear for earlier-type stars. The purpose of a primarily automated search method is to work through vast amounts of data with a systematic approach and test different search parameters while reducing human error. Section \ref{sec: data} describes the TESS data, Section \ref{Method} discusses our updated search method, Section \ref{sec: results} explains the results and filtering process of false positives, Section \ref{sec: candidates} shows our candidates from the search,  we discuss the occurrence rates in Section \ref{sec: discussion}, we set a future outlook in Section \ref{sec: outlook}, and conclude in Section \ref{sec: conclusions}. 


\section{Data} 
\label{sec: data}
The TESS light curves used in this work are the \verb|eleanor-lite| release of the \verb|eleanor| light curves \citep{FeinsteinEleanor, updated_eleanor_lcs2022}, downloaded from the Mikulski Archive for Space Telescopes (MAST). This pipeline generates 30-minute cadence light curves for all stars brighter than 16th magnitude in TESS' $T$ band, making it the largest TESS dataset publicly available to date. The light curve data product for a given star contains three flux time series columns: the Raw Flux, Corrected Flux, and the PCA flux. The units of time in these data are Barycentric TESS Julian Date (BTJD), which is ${\rm Julian~Date (JD)} - 2457000$. The Raw Flux is the background-subtracted photometry extracted by multiplying the aperture mask and the Target Pixel Files (TPF), which is a 3D cube of images over time, and summing the pixels in each cadence. The Corrected Flux improves on the Raw Flux by correcting for systematics on an orbit-by-orbit basis. The PCA (Principal Component Analysis) flux uses PCA to further remove systematics from the flux values that might stem from neighbouring stars on the same detector. We use the PCA flux data product, which is created by subtracting the first 3 cotrending basis vectors (CBVs) for the appropriate camera to further reduce variability. 

Each of the light curves contains a series of quality flags for each cadence. Quality flags are used to note any events that may cause variations in the flux that were picked up by the pipeline, such as cosmic rays, argabrightening events (sudden brightening across the CCD cameras), attitude tweaks, and reaction wheel zero crossings, which are all marked with a non-zero flag. We also created a new flag for all data points that exceed our sigma-clipped ensemble noise threshold, corresponding to ramps in the data as a result of scattered light from the Earth and Moon as the spacecraft moves on its orbit  (see Section \ref{subsec: removing common systematics}). The light curves also contain background flux and centroiding columns, which are used to identify false-positive events from short-term fluctuations that occur around the same time as an event.

It is important to note that the aperture selection for saturated targets ($T < 6.8$) in \verb|eleanor-lite| does not capture all flux: overcharged pixels for bright targets could bleed onto other pixels, and the pipeline does not automatically correct the aperture for every target. Therefore, to address potential issues, we additionally used light curves from the TESS-SPOC pipeline \citep{spoc} for stars with $T < 6.8$. The TESS-SPOC pipeline is a smaller sample compared to \verb|eleanor-lite|, with up to 160,000 targets per sector, and selects targets based on certain criteria optimised for exoplanet searches around M-dwarfs (bright targets in the near-infrared, within 100 pc, and $T \leq 13.5$), whereas \verb|eleanor-lite| provide photometry for all stars on the detector. The TESS-SPOC light curves were downloaded from MAST, where we used the Presearch Data Conditioning SAP (PDCSAP; \citealt{Stumpe2012,Stumpe2014,Smith2012}) data products, which correct for systematic trends shared with other stars on the detector (co-trending basis vectors) and is corrected for dilution. Our total sample covers the entire TESS primary mission, sectors 1-26, for stars with a magnitude of $T < 13$. We apply this magnitude cutoff based on the estimated number of star-years observed and injection-recovery tests, which we discuss in Sections \ref{subsec: choice of sample} and \ref{sec: injection recovery}.
\\

\subsection{Choice of Sample}
\label{subsec: choice of sample}
As exocomets are expected to produce low-amplitude transits of a sub 1\% order, there will be a point at which trying to detect them would be unlikely due to the noise properties of the light curves. Compared to \textit{Kepler}, TESS has smaller optics and is in an elliptical high-Earth orbit, introducing scattered light and hence has greater noise for a star of a given magnitude.


Conducting a large-scale search for exocomets would only be likely to be successful compared to \textit{Kepler} if the chosen TESS sample was comparable. \textit{Kepler} observed approximately 150,000 stars continuously over its 4-year mission, a cumulative observation time of roughly 600,000 star-years. We consider the concept of star-years valuable here because (except for $\beta$~Pic) photometric exocomet transits are rare; the probability of observing a transit should therefore be proportional to the observing duration, and observing 100,000 stars for 1 year approximately equivalent to observing 200,000 stars for 6 months. Because the noise properties are primarily a function of stellar magnitude, we aim to make a magnitude cut in the TESS sample such that we have a similar or greater number of star-years as \textit{Kepler}. Given the 27d length of each TESS sector, roughly 8.1 million TESS light curves (or a magnitude cut of $\sim T = 12.5$) are needed for a comparative 600,000 cumulative years of observation which should, like \textit{Kepler}, yield a few detections (i.e. about 1 detection in 200,000 star years). TESS light curves are approximately 27 days long for each sector. We round our sample size to an upper magnitude limit of Tmag = 13, where we have $\sim$ 15 million light curves, corresponding to $\sim$ 13 million stars. The magnitude cut at $T = 13$ provides a noise level of no more than 1000 ppm \citep{FeinsteinEleanor}, which is similar to the level of the expected exocomet transits. Including fainter stars will therefore not necessarily yield more detections, which we explore quantitatively with injection recovery tests in Section \ref{sec: injection recovery}, but aligns with the results of our injection-recovery tests, where our sensitivity for comet transits injected at a 0.1\% depth falls sharply beyond $T = 13$. TESS is therefore expected to be comparable to \textit{Kepler} for exocomet searches, with the advantage that this calculation is based on only the first two years of data from TESS.



\section{Method}
\label{Method} 
The search method framework is based on the work in \cite{Kennedy2019}, where single transiting events in a light curve are identified based on computing a test statistic and calculating its signal-to-noise ratio, followed by fitting symmetric and asymmetric Gaussian models and comparing their residuals to quantify an asymmetry value. Given the nature of the TESS mission and reflection on the \textit{Kepler} work, changes were required. The details of the method and the updates are described below.

\subsection{Removing Common Systematics}
\label{subsec: removing common systematics}
\begin{figure*}
\centering{\includegraphics[scale=0.5]{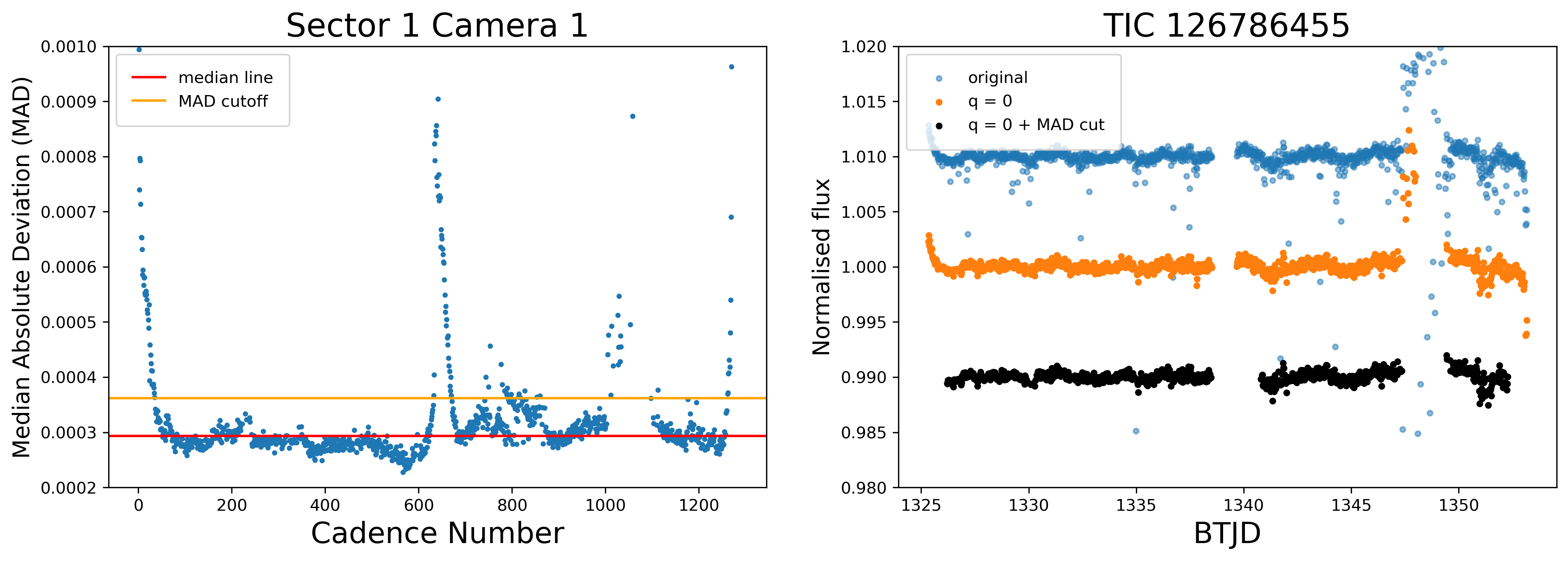}}
\caption{An example of applying the MAD filter to a light curve. The left plot shows the MAD(t) values across the Sector 1, Camera 1 combination, where the red line is the median MAD(t) value, and the yellow line represents the MAD threshold. The right plot shows the light curve of TIC~126786455, where the blue line is the original light curve, the orange light curve removes the non-zero quality flags only, and the black light curve removes non-zero quality flags and applies the MAD cut. The MAD cut removes data near the beginning and end of the light curves, and the bad data points around BTJD 1348. The light curves are offset for visual purposes.}
\label{fig: MAD}
\end{figure*}

The first step is to clean the light curves of all known systematics by discarding all data points with non-zero quality flags. Beyond the events picked up by the quality flags, there are systematic effects at the edges of light curves and data gaps mainly due to scattered light as a result of the location of the spacecraft. An example of this can be seen with the raw light curve in Fig \ref{fig: MAD}. This can introduce unwelcome false positives, where ramps can be misclassified as an event. To remove the ramps and other flux variations, we used the 30-minute cadence light curves for the 2-minute cadence targets in a given sector-camera combination to compute the median absolute deviation (MAD) for each cadence. The MAD is given by the median of the set of absolute deviations from the median flux at that time:
 
\begin{equation}
    MAD(t) = median(\{|F_{lc}(t)-\bar{F}(t)|\}),
	\label{eq:MAD}
\end{equation}

where $F_{lc}(t)$ is the normalised flux of a light curve at time t, $\bar{F}(t)$ is the median flux of all normalised light curves in the set at time t. After calculating the MAD for each sector-camera (i.e. the variation in MAD(t) for each sector), we also compute the median MAD for each sector-camera combination. Cadences more than 3 $\times$ from the median MAD value are assigned a new quality flag 65536, so cadences with this flag assigned are removed along with all other non-zero flags. An example of the MAD calculations and their application to a light curve is shown in Fig \ref{fig: MAD}, where the common systematics at the edges of the light curves and near the gaps are removed while preserving most of the data. 

\subsection{Choice of Smoothing Filter}
\label{subsec: choice of smoothing filter}
It is expected that many of the TESS light curves will show some form of astrophysical variability, particularly with the young stars in this sample where exocomets might be more likely to be detected. This includes periodic rotational modulation and stellar pulsations. Although one can remove this variability for a small sample of stars more thoroughly, large-scale searches require a more generic approach. \cite{Kennedy2019} used a Lomb-Scargle periodogram to remove astrophysical variability, where peaks above a certain threshold were identified, followed by iterative subtraction of the corresponding sinusoids from the flux. This worked well with the long-cadence \textit{Kepler} light curves due to the 3-month long observation period per light curve. However, we discovered that when reimplementing this technique with the TESS light curves, the 30-minute cadence data coupled with the 27-day TESS observation window proved to be too small a number of points per light curve for this method to work as expected compared to the longer \textit{Kepler} light curves or higher cadence TESS data. In particular, in our early tests, we tried the Lomb-Scargle removal technique on the $\beta$~Pic exocomet detection from \cite{Zieba2019} to see whether the transit could be successfully recovered. However, the exocomet transit itself was removed from the light curve. With the 2-minute SPOC light curve of $\beta$~Pic available, the removal of sinusoids worked as expected, and the exocomet transit was recovered successfully. This meant that for the case of $\beta$~Pic and its $\delta$~Scuti pulsations (period of $\sim$ 20-60 minutes) the difference in cadence used to observe the star could not adequately separate the pulsations from the exocomet event. As we are working with 30-minute light curves and are analysing them individually (i.e. targets with multiple observations were not analysed together), where we would have to consider a wide sample of stars, we opted for non-frequency-based alternatives to detrend light curves. 

Multiple detrending techniques were experimented with, including low-order polynomial fitting \citep[LOWESS;][]{Cleveland1979}, cadence-based windows \citep[e.g.: Savitzky-Golay filters;][]{savgol1964} and time-based windows (mean/median filters). We concluded that a sliding median filter with a 2.5\, day window from \verb|wotan| \citep{hippke_wotan} was adequate to preserve cometary shapes (which were based on the \textit{Kepler} candidates and $\beta$~Pic) while remaining computationally fast. The chosen window size was decided from the upper limit of current exocomet candidates. 

\subsection{Identifying Transit Events}
\label{subsec: identifying transit events}
The search method consists of finding single transits, where the most significant drop in flux in a light curve is identified by a moving average over a range of window lengths. The light curve consists of time and normalised flux values. To account for unevenly spaced data, such as the day-long gaps, data are linearly interpolated to fill all gaps to make the steps below simpler in practical terms, but flagged as ``fake'' where the interpolation spans more than 3 cadences (1.5 hours). We discard transits within 1.5 days of a large data gap, so this interpolation does not affect the results. We use a test statistic as in \cite{Kennedy2019} to identify the most significant transit event in a light curve:







\begin{equation}
    T_{m,n} = \frac{(\bar{x}_{m,n} - 1)\sqrt{2m}}{\sigma_x}, \qquad  \bar{x}_{m,n} = \frac{1}{2m} \sum\limits^{n+m}_{i = n-m+1} x_i.
\end{equation}

\begin{figure*}
\centering{\includegraphics[scale=0.57]{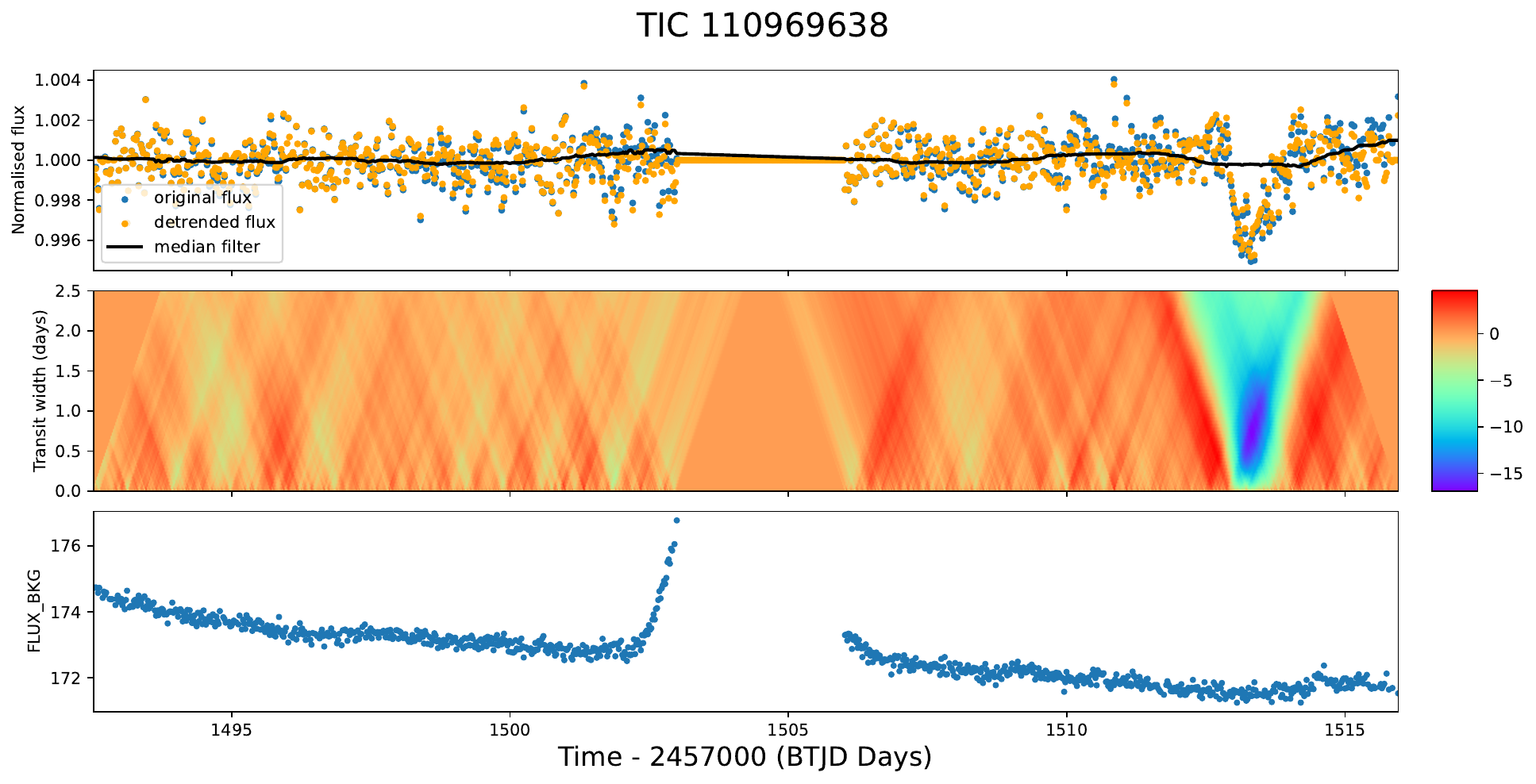}}
\caption{The light curve of TIC~110969638 (top panel), its corresponding test-statistic $T_{m,n}$, (middle panel - the colourbar shows the $T$ value across the light curve), and the background flux activity of that light curve (lower panel). There is an evident transit event at BTJD 1513 with a $T_{min}$ = -16.92, corresponding to an SNR = 6.43 with equation \ref{eq: SNR}. Changes in the background flux that correlate with a transit could indicate that the transit is a false-positive, but there is no indication of that here. }
\label{fig: search-method}
\end{figure*}

Here, the light curve is a vector $x$ of length N with RMS $\sigma_x$ (calculated as the standard deviation of the normalised flux) measured for each light curve.  The window length is 2m points wide. The statistic is therefore a moving average at each point in the light curve $n$ that are valid at a given half-width $m$. The widths ranged from 0.1 - 2.5 days wide, which is the same as \cite{Kennedy2019}, inspired by the exocomet transit duration times from \cite{Rappaport2018}, and given the transits seen for $\beta$~Pic, covers the range of known transit durations. The deepest event is the most negative $T$ value in the light curve at indices $m$, $n$. While $T$ is an estimate of the signal to noise ratio (SNR) of a given transit, astrophysical or other variation means that it can be overestimated. Therefore, after the most negative $T_{m,n}$ is found, we calculate the SNR by dividing T by the RMS of the T-statistic,


\begin{equation}
    SNR = - min(T_{m,n}) / \sigma_{m},
    \label{eq: SNR}
\end{equation}

where $\sigma_{m}$ is the standard deviation of $T_{m}$ at width $2m$. This $\sigma_{m}$ is a slight modification compared to the standard deviation used in \citep{Kennedy2019}, which was computed over all $m,n$. While the search method can identify multiple transiting events, we restrict the algorithm to only focus on the most significant negative event in the light curve, as multiple detections could introduce many false positives, and identifying multiple exocomets in a system as part of the automated procedure is beyond the scope of the current work. Fig \ref{fig: search-method} shows a plot of this search method. 


\subsection{Quantifying the Transit Shape}
\label{subsec: Defining the transit shape}
Since the T-statistic identifies an event without characterising its shape, results from the search will yield many events that do not have a comet-like profile, including exoplanet transits and eclipsing binaries. False positives could still arise from the search, such as systematic artefacts with a significant drop in flux such that they appear transit-like. To characterise these events, we fit two functions for each event: a Gaussian profile which imitates a planet/binary transit due to its symmetry, and a skewed Gaussian profile. In \cite{Kennedy2019}, a Gaussian profile and an asymmetric Gaussian model (made from a half-Gaussian, half-exponential from the mid-transit) were fit and their residuals were compared to provide an asymmetry ratio, $\alpha$. Although this worked well, we found that there were some limitations in the definition of the asymmetry parameter, the main one being that the asymmetric Gaussian model is a combination of two functions, which assumes that we physically assume an exocomet-like transit. Using a model that accounts for asymmetry more generally makes it easier to rule out false positives that are asymmetric in the opposite direction, mainly attributed to systematics or unsubtracted stellar variability. 





To address this issue, we replaced the asymmetric Gaussian with a skewed Gaussian. The skewed Gaussian model makes use of the skew-normal distribution, a continuous probability distribution that generalises the normal distribution to allow for nonzero skewness. The symmetric Gaussian model is given by:

\begin{equation}
   \psi(A,\sigma,t_{\rm tr},t) = A e^{-\frac{1}{2}(\frac{t-t_{\rm tr}}{\sigma})^2},
\end{equation}

where $A$ is the amplitude, $\sigma$ the width, and $t_{\rm tr}$ the transit centre, and time $t$. $x_{\rm sym}$ is then defined as:

\begin{equation}
    x_{\rm sym}(A,\sigma,t_{\rm tr},t) = 1 - \psi(A,\sigma, t_{\rm tr}, t).
\end{equation}

For the skewed Gaussian, we need the probability density function $\phi(x) = \psi \left(\frac{1}{\sqrt{2\pi}}, 1, 0, x \right)$:

\begin{equation}
    \phi(x) = \frac{1}{\sqrt{2\pi}}e^{-\frac{x^2}{2}},
\end{equation}

and its cumulative distribution function:

\begin{equation}
    \Phi(\beta) = \int_{-\infty}^{\beta} \phi(t) dt = \frac{1}{2} \left (1 + {\rm erf}(\frac{\beta}{\sqrt{2}}) \right ),
\end{equation}

where erf is the error function. Our model is based on the probability density function for a skew-normal distribution with a skewness parameter $\beta$, and is given by:

\begin{equation} \label{skewed gaussian}
    x_{\text{skew}} = A \left( 2\phi(t-t_{\text{tr}})\Phi(\beta(t-t_{\text{tr}})) \right).
\end{equation}

The pure Gaussian model was fit first, and the skewed Gaussian second, using initial parameters from the pure Gaussian fit and $\beta = 0$. A symmetric transit should then result in the same fit in both cases since non-zero $\beta$ would not improve the fit. The parameters for this model were optimised using the \verb|scipy curve_fit| function, which uses a least squares approach \citep{scipy}. 


The model fit was applied to a cutout of the light curve data centred on the mid-transit point $t_{tr}$ computed from the location of the minimum $T_{m,n}$. The cutout was determined to be 3 transit widths on either side of the centred time, where the transit widths were the widths computed from the T-statistic to be the optimal transit width for the strongest signal. Due to the shorter light curves with TESS, choosing the number of transit widths was important as more would rule out potential candidates that were detected close to the edges of the light curve data. In addition, a linear trend was subtracted from the cutout using the data within three transit points of the start and the end to remove any local trends which were not removed by the smoothing filter. This linear subtraction was also necessary because a symmetric dimming event that is superposed on a decreasing linear trend would also appear asymmetric, potentially creating a false positive candidate \citep{Kennedy2019}. With the above changes in replacing the two-part asymmetric Gaussian model with a skewed Gaussian model from a continuous probability distribution, the redefined asymmetry ratio $\alpha$ is then:


\begin{equation}
    \alpha = \frac{\Sigma(x_i - x_{sym})^2}{\Sigma(x_i - x_{skew})^2}
\end{equation}

where the sum is taken over the light curve cutout. This model ratio determines which of the two models fits the transit event more accurately. An $\alpha=1$ implies that both the symmetric Gaussian and skewed Gaussian fit the event equally well, implying the transit event corresponds to a symmetric profile. Therefore, events such as eclipsing binaries, exoplanet transits, and other symmetric dips would have an asymmetry value of or very close to 1. An $\alpha > 1$ would indicate that the asymmetric model fits the data better. Using the skewness value $\beta$ of the skewed Gaussian model complements the asymmetry ratio, as we can then determine the direction of the asymmetric transit and how asymmetric the transit event is. A positive $\beta$ would be the comet-like transit shape that we are seeking, and a negative $\beta$ would represent an event that has a gradual ingress followed by a steep egress. From a visual inspection of negative $\beta$ values, this is common with cases of stellar variability where the search method has identified the most significant event to be one of the oscillations of the rotational modulations, such as those that remained after detrending.

\subsection{Injection Recovery and Testing}
\label{sec: injection recovery}
To provide a quantitative assessment of our recovery rate, and to find the magnitude limit at which stars are too faint to recover exocomet transits with our search method, we perform injection and recovery tests with a random sample of light curves of each magnitude. 

To accurately reproduce light curve characteristics, such as the data gaps and noise properties at each magnitude range, we used real light curve data from the \texttt{eleanor-lite} sample mentioned in Section \ref{sec: data}. We randomly selected 20,000 light curves in each magnitude range from $T = 3$ to $T = 15$ to inject transits. For the brighter magnitudes where there are fewer light curves than the sample size, light curves can be selected multiple times. We injected synthetic comet transits using the skewed Gaussian model with the initial parameters based on the \cite{Zieba2019} $\beta$~Pictoris exocomet detection at random depths and times in the light curve. The range of depths of exocomet transits was chosen to be from 0.01\% to 1\% to cover a wide dynamic range. The transits are injected and then follow the search method procedure in Section \ref{subsec: removing common systematics} - \ref{subsec: Defining the transit shape} with cuts that are the same as those used to search for exocomets below. The criteria for a recovered injection is for the comet transit to be recovered within a 5-hour window from the injected time. The depth is allowed to be different to the injected depth, as the recovered depth can be very different in noisy light curves.

 

\begin{figure*}
\centering{\includegraphics[scale=0.39]{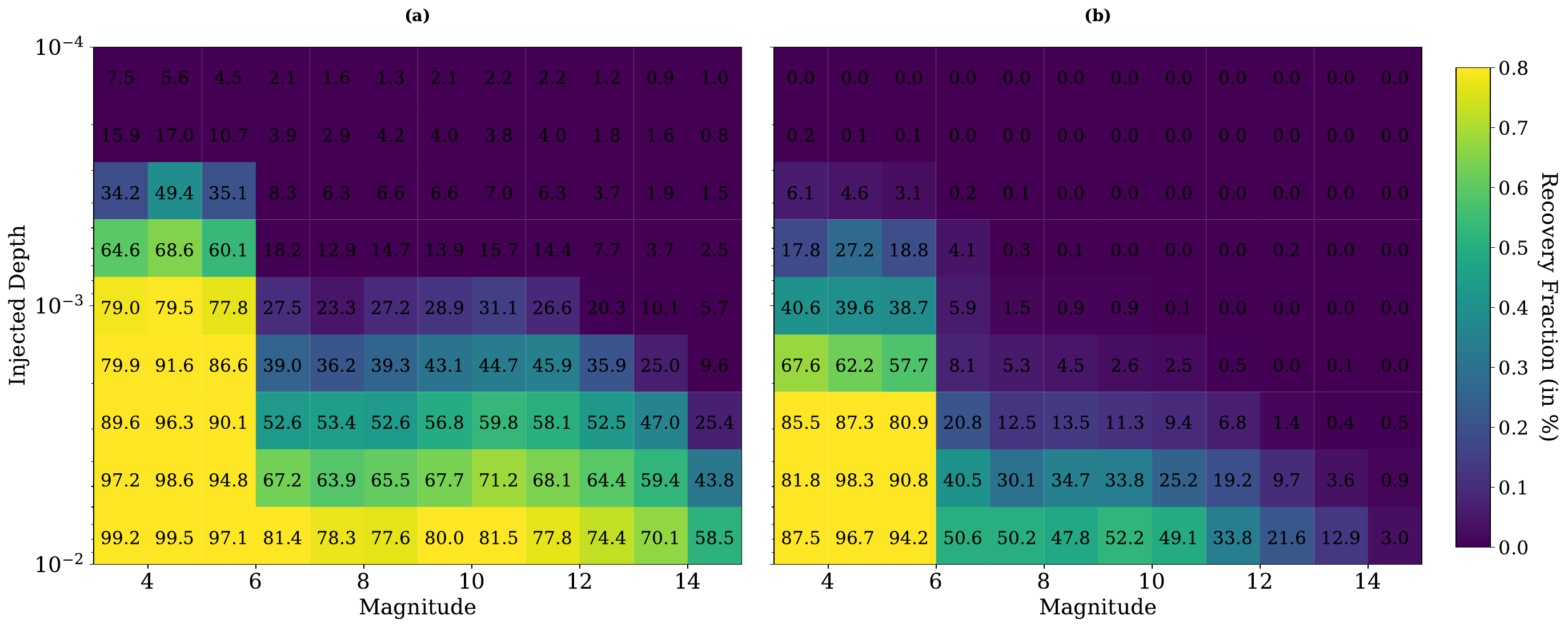}}
\caption{Two-dimensional detection efficiency maps for our search method across the magnitude and depth range. Each magnitude bin consists of 20,000 injected transits. The sensitivities in plot (a) take into account the transit search if the injected transit was recovered within a 5-hour window of the injection time, and plot (b) uses the criteria for plot (a) \textbf{and} our thresholds applied in the analysis steps of the pipeline.}
\label{fig:injection_recovery}
\end{figure*}

The results of the injection recovery tests are shown in Fig \ref{fig:injection_recovery}, which shows the fraction of injected exocomet transits recovered over a range of depths and magnitudes. The left panel (a) shows the raw recovery rate of the search method, set only by the requirement of a transit within 5 hours of the one that was injected. The right panel (b) shows the real recovery rate, which includes the additional cuts used for the real search that are described in Section \ref{sec: results}. The plots show that the raw recovery rates are at least tens of percent across the grid down to transit depths of 0.1\% and up to $T = 13$. However, the real recovery rate is much lower than the raw rates, mainly because our ability to characterise the transit shapes drops significantly below depths of 0.3\%. There is also a significant drop in the recovery rate beyond 13th magnitude, from 20.31\% to 10\% between $T = 13$ and $ T = 14$ at 0.1\% depth. Taking into account the need for a large sample size to compare our observations with \textit{Kepler} (see Section \ref{subsec: choice of sample}), and the drop between 13th to 14th magnitude in recovery rate, we thus set $T=13$ as our magnitude limit for our sample of stars. The results from these tests and \ref{subsec: choice of sample} show that getting to 0.1\% level transits will be challenging for most stars as the recovery rates are only a few percent.  

\section{Results}
\label{sec: results}

The search was performed on a sample of 15,477,322 30-minute cadence TESS light curves (9,224,280 unique TIC IDs), and we performed several stages of rejection to remove false-positive events. Table \ref{table:candidates filtered} shows the number of candidates remaining at each vetting stage. The first pass rejects events with insufficient data, events too close to the data gaps, and events where the fits did not converge. Of the 15.5 million light curves, 6.4 million light curves yielded an event where transit profiles were successfully fitted. We then examined the distributions of several parameters to get some threshold-based cuts. We initially experimented with Sector 6 (which includes $\beta$~Pic's deep transit), and then included Sectors 9 and 17 to check that the choices were not biased by some characteristic of the Sector 6 data. The first threshold was the SNR, where we followed the cut made in \cite{Kennedy2019} and cut all transits with SNR < 5 to ensure that the event was significant in the light curve. Of the 6.4 million light curves, 553,294 events corresponded to SNR $ > 5$. 


\begin{table}
  \centering
  \caption{Number of candidates remaining after each stage of vetting.}
  \label{table:candidates filtered}
  \begin{tabular}{|r|c|}
    \hline
    No. of targets & Cut applied \\
    \hline
    15,477,322 & None (full sample) \\
    6,424,883 & Data-based cuts \\ %
    553,294 & SNR > 5 \\ 
    184,828 & Duration > 0.4d \\ 
    123,819 & Depth < 1\% \\ 
    8,358 & Common time transit cut (see \ref{Common time transits})\\
    3,515 & 0 < Skewness < 29 \\
    910 & Asymmetry > 1.02 \\
    272 & Skewness < 8 \\
    12 & Visual vetting \\
    6 & Final Vetting \\
    \hline
 \end{tabular}
\end{table}



\begin{figure}
\centering{\includegraphics[scale=0.4]{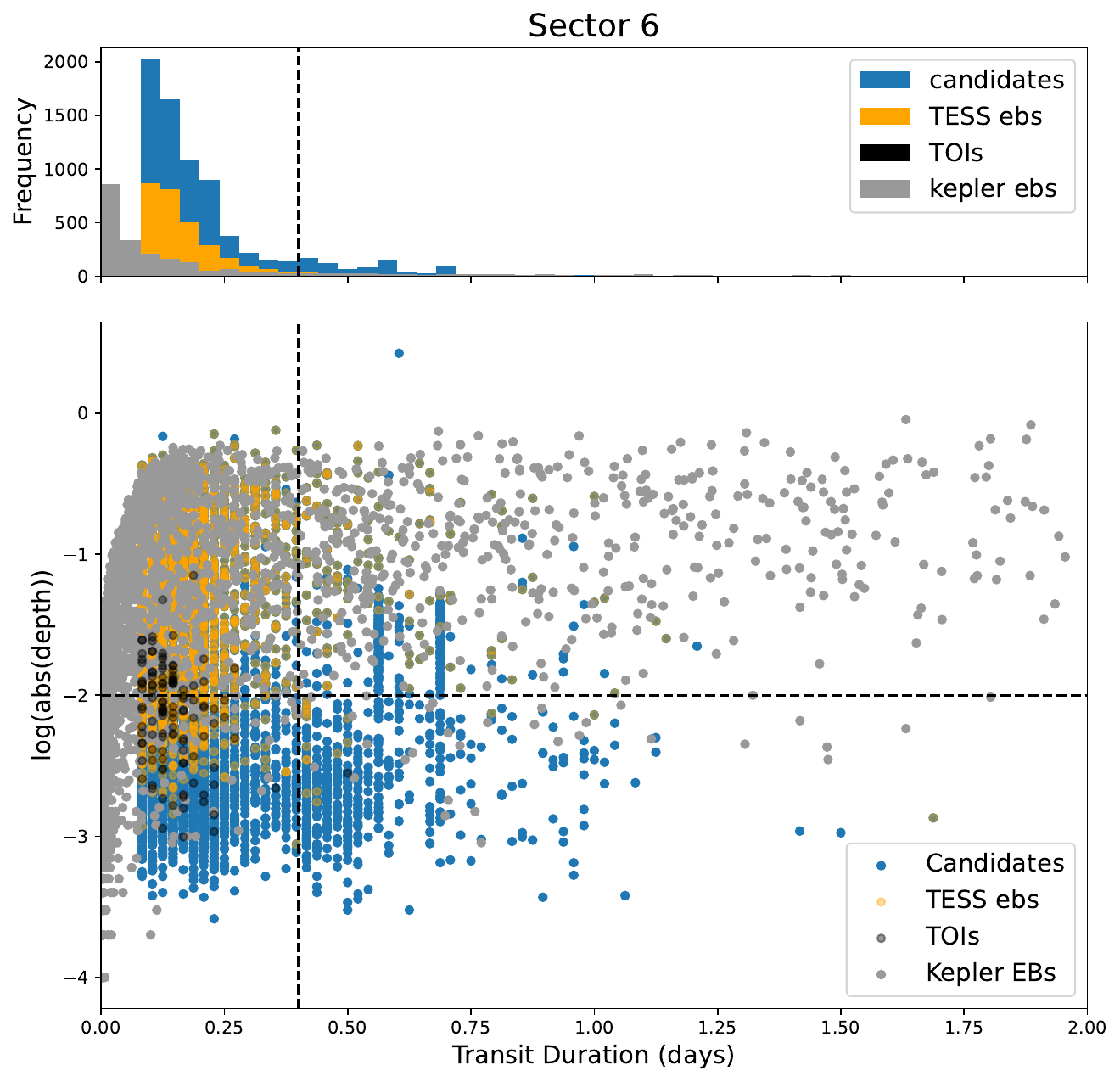}}
\caption{Distribution of the candidates in Sector 6 showing the transit duration against the log transit depth. The candidates are coloured in blue, transits from the TESS eclipsing binary catalogue in \citep{eb_catalog_ethankruse} are coloured in orange, TOI transits from EXO-FOP \citep{exofop} are coloured in black, and transits from the \textit{Kepler} Eclipsing Binary catalogue \citep{kepler_eb_catalog} are coloured in grey. \textit{Kepler} did \textbf{not} observe the equivalent of TESS sector 6, we are only overlaying this catalogue to demonstrate justifying our duration cut. For the known sources, most objects lie before the 0.4d duration cut. Known sources that lie beyond the 0.4-day cut are less than 3\% per object type, but they are also removed from our remaining sample. The vertical dashed line therefore shows the 0.4d cut, and the transit depth threshold cut at 1\% is shown as the horizontal dashed line.}
\label{fig: s6-duration-cut}
\end{figure}

The main distinction between exoplanets and exocomets is their transit durations, where exocomets have been observed to have longer durations. The duration of the largest exocomet transit around $\beta$~Pic was $\sim$ 2 days from \cite{Zieba2019}, and similar transit durations were also seen in the \textit{Kepler} candidates from \cite{Rappaport2018,Kennedy2019}. To estimate a practical lower limit, we compared the distribution of durations from the T-statistic to durations from the following catalogues: the \textit{Kepler} Eclipsing Binary Catalog \citep{kepler_eb_catalog}, the ExoFOP-TESS list of TOIs \citep{exofop}, and the TESS Eclipsing Binary catalogue created from the \verb|eleanor| pipeline \citep{eb_catalog_ethankruse}. We found that most of the TOIs and binary sources lie at transit durations shorter than 0.4 days. For the TESS EBs, only 4.8\% of the catalogue existed beyond this cut. This can be seen in Fig \ref{fig: s6-duration-cut}. Given what is known about the \textit{Kepler} binaries also seen in Fig \ref{fig: s6-duration-cut}, and the rarity of exocomets and their transit durations, it is more likely that a transit duration shorter than 0.4d is an eclipsing binary than an exocomet. We therefore assume that most transits of duration $<$ 0.4 days are more likely due to binaries than exocomets, and discarded all candidates with transit durations shorter than this threshold. We also removed all known TOI and eclipsing binary sources from our sample, including those that remained beyond this cut. After this cutoff, we were left with 184,828 candidates. 


The last major threshold we derived from our distributions is the depth of the transit events. The deepest exocomet transit discovered to date in TESS from \cite{Zieba2019} only has a depth of 0.2\%. The sample size of exocomet discoveries in photometry is still very small, and it is still unknown what a generic expectation for transit depths should be. We cannot rule out that other systems may host deeper exocomet transits, but we may expect that the deepest events would be observed first. However, the caveat is that deeper transits are likely to be rarer than shallower ones \citep{lecavalierdeseetangs2022}, and that $\beta$~Pic itself may have deeper transits that we are yet to detect. Therefore, we set an arbitrary practical threshold of a 1\% transit depth, where all transits deeper than this are excluded. After this cut, 123,819 candidates remain. 

\subsection{Common time transits}
\label{Common time transits}
An unexpected false-positive that emerged from our remaining 123,819 candidates was ``common time transits'', where clusters of transit events occurred around the same time in an unusually large number of light curves, with no underlying relationship between the targets. Using a subsample of these artefacts, we compared these light curves with other pipelines where those light curves were available and found that they were not present in those pipelines. These events were also seen in parallel work in Cody et al. (in prep.), who concluded that these were processing issues with the light curves. We decided to remove all transits that were detected in these common times, which in some sectors dominate the candidates. Although this can remove some real transiting events, we acknowledge this as a necessary step to proceed with our candidate search. We removed common-time transits by binning our transit times into 100 bins per sector and set a threshold of 10 times the MAD of the bin counts, where all bins containing counts more than this threshold were removed from the sample. Applying this across all our 26 sectors cut down our sample size to 8,358 candidates.

\subsection{Characterising transit shape}
We then characterised the events based on shape properties derived from the skewed Gaussian model. The shape of an exocomet transit, a sharp ingress followed by exponential-like egress, corresponds to a positive skewness value. A negative skewness would imply that the coma is optically thinner and longer than the tail, which is not expected by the transit shape predictions \citep{Lecavalierdesetangs1999}. We cut all candidates with a negative skewness value, leaving us with a sample of 3,515 candidates. 


For the remaining candidates, we had initially experimented with Self-Organising Maps \citep[SOM,][]{kohonen_som} to cluster candidates based on their transit shape. The idea was that distinct shapes would emerge from the dataset which can pair symmetric transits with others on a ``Kohonen grid'' - a two-dimensional map where similar transit patterns would cluster together in neighbouring regions - and that asymmetric transit shapes would appear as their own cluster. However, we found that the SOM could not accurately match the shapes together due to an imbalanced dataset, where asymmetric shapes were merging with the more characterised symmetric clusters. Therefore, we used alternative methods to exclude false positives, and decided to create two more cuts. The first was related to the asymmetry ratio. By inspecting a subset of the remaining sample at various $\alpha$ values, we create a threshold such that only light curves with an asymmetry above 1.02 are deemed to have an asymmetry ratio that can be distinguished from a symmetric transit (i.e. a slightly more generous cut than the 1.05 used in \cite{Kennedy2019}). In addition, an inspection of candidates showed that light curves with skewness values $\geq 8$ represented discontinuities in the flux, or poor fits to the data, mainly attributed to noise, and hence an inability to quantify any asymmetry robustly. We examined the skewness parameter space by taking subsets of data and visually assessed where models fitted to discontinuities rather than ``true'' events. We set an upper skewness threshold of 8, and also exclude candidates where the skewness error exceeds 50\% of the skewness value. We were left with 272 high-quality transit candidates.



\subsection{Manual vetting}
The remaining 272 candidates were visually inspected. The first step was to check the background fluxes and centroids for each target from the FITS files. These columns in the data products can identify false-positive events such as asteroid crossings or deeper transits from nearby stars whose PSF overlaps with the target. Transit events coincident with increases in the sky background are mainly caused by nearby asteroids passing through background pixels and were therefore excluded. We then relaxed the quality flags to include flagged data and inspected the RAW FLUX of the light curves to ensure that there was not an issue with the computed PCA FLUX. We examined the light curves in other TESS sectors where available to determine whether the observed event was periodic. As \verb|eleanor-lite| produces light curves for the primary mission only, we used TESS-SPOC and the Quick-Look Pipeline (QLP) \citep{Huang2020_QLP1,Huang2020_QLP2} pipelines for light curves in the extended missions. We removed signals where the strongest events were attributed to binaries, stellar pulsations, and tidal interactions.



The TESS pixel size is such that the apertures used to extract photometry for many of our candidates contain flux from neighbouring stars. This problem is caused in part by the coarse pixel scale of TESS (21"/pixel) and is amplified near the galactic plane where the density of visible stars increases significantly. For our candidates, we calculated the true transit depth using:
\begin{equation} \label{eq: dilution}
    f_{\rm real} = f_{\rm obs}  \left(1 + \frac{F_B}{F_A} \right), 
\end{equation}
where $F_A$ is the flux of the target star, $F_B$ is the flux from the neighbouring star, $f_{\rm obs}$ is the observed fractional transit depth, and $f_{\rm real}$ is the corrected transit depth. For $F_B$, we selected only the brightest star closest to the target, where the star would be the most likely to contaminate the transit. Once an estimate for the corrected transit depth had been calculated, we removed candidates that exceeded the 1\% transit depth threshold. After inspecting the 272 candidates and rejecting candidates based on their background, periodicity checks and dilution, 12 candidates were left for further inspection. 


We found that 6 of the remaining 12 candidates were false positives that required a detailed visual inspection to verify. One false positive that passed through all the cuts was $\eta^1$ Pic (TIC 219367750, or HD 50241), a Tmag = 4.98, F5V star where a 0.14\% transit was detected with an SNR of 5.11 at BTJD 1441. The system has an age of roughly 1400Myr, with a distance of 26.2 pc. This star has strong rotational modulation, much shorter than our 2.5-day smoothing window; however, with more aggressive detrending with both shorter smoothing windows for our median filter and the Lomb-Scargle subtraction method, a signal persisted. This was interesting as $\alpha$~Pic is also a candidate exocomet host star in spectroscopy \citep{Hempel2003}, and it was unexpected that a star close to $\beta$~Pic would be one of the few candidates out of our 15.5 million light curve sample to appear as a candidate. Exploring this further, we followed the steps introduced in \cite{lecavalierdeseetangs2022}, where we inspected bright nearby stars and applied the thorough detrending procedure with the Lomb-Scargle iterative subtraction method to those light curves. We downloaded the 2-minute SPOC light curves for the bright stars in the Pictor constellation: $\alpha$~Pic, $\eta^2$~Pic, $\gamma$~Pic, $\theta$~Pic, and $\zeta$~Pic, as well as the $\beta$~Pic light curves. There were no dimming events detected with the associated stars at BTJD 1441. However, we later found potential systematics associated with this event in the 2-minute cadence data, though the 30-minute light curves did not show an obvious discontinuity due to its binning of the 2-minute data. A consistent scatter in both the X and Y-centroiding at the transit times was present in all those stars that were only present in the 2-minute light curves, seen in Appendix \ref{Appendix: ruling out alpha pic}. This could be one of the reasons there are issues in the $\eta^1$ Pic light curves, where overcharged pixels for a saturated target could lead to poorly resulting light curves, and we do not consider the target further. 

In addition, we found that three of the remaining candidates were false positives when examining their PSF\_FLUX available with the \verb|eleanor| user interface. The PSF\_FLUX helps in separating the flux from nearby stars on the TESS detector. CHSS 93, TIC~131422529 and TIC~349766841 were candidates that had several nearby stars in their TESS TPF, and upon inspecting their PSF\_FLUX and pixel-by-pixel light curves, found that the transit event for these systems were not significant enough to be deemed convincing candidates, and we do not consider them beyond this point. TIC~119918912 was ruled out as it appeared to have a single transit in its light curve, but was found to be periodic when the data from other TESS sectors were inspected. The other light curves of TIC~119918912 showed dimming events with an orbital period of approximately 20d, where the transit shapes also appeared to be asymmetric, which would have been interesting, but upon inspection are more likely attributed to linear trends. Gaia identified this target as an SB1 binary, where the NSS solution suggests the transit is from the binary itself and not an exocomet event. The last false positive candidate was TIC~419365265, where the dimming event appeared to originate from the neighbouring star rather than the target as the transit depth for the nearby star was deeper than the target star at the corresponding time. 

We note that in our vetting, we did not capture any obvious dipper stars such as those in \cite{capistrant-rappaport-dippers-2022,Gaidos2022,tajiri-dippers-tess-cnn-2020}, which can show large variations thought to be related to stellar occultation from parts of a circumstellar disc. This is most likely due to the depth threshold that we have set for our sample, or that the multiple dips in a dipper star light curve reduce the overall SNR of the T-statistic. While the depths vary for dippers, their most significant events are more likely to be deeper than 1\%, which are removed in our cuts. Expanding the search for dippers is beyond the scope of this work, but an interesting one to explore given the proposed origins of the circumstellar dust observed in these systems potentially being from cometary infall \citep{ansdell2019}. 

\subsection{A complementary search with the full TESS-SPOC data}
While the primary motivation to use the \verb|eleanor-lite| light curves was its large, magnitude-limited sample of targets, the limitations faced such as common-time transits may have potentially missed some real transit events. A complementary search of the TESS data was performed with the TESS-SPOC light curves, which include stars down to $T \leq 13.5$ as a check for that subset of stars (bright targets in the near-infrared and within 100 pc) that may have been removed by a common time cut during the vetting stage with the \verb|eleanor-lite| light curves or were bright targets that were not optimally processed due to the saturation limitations of the \verb|eleanor-lite| pipeline. We applied the same cuts as above to manually inspect the remaining candidates. After applying identical cuts from the search above, 250 candidates were manually vetted. From our complementary TESS-SPOC search, only $\beta$~Pic (TIC 270577175) and TIC 280832588 were available in the pipeline sample, and no new candidates were identified. 



\subsection{A search for `reverse' exocomets} \label{diagnostic check}

We also performed a time-reversed search to test whether transit-like events with reverse asymmetry are common in the TESS data. There might be astrophysical or instrumental phenomena that cause asymmetric transits that look like either `forward' or `reverse' exocomets; this search could, in principle, provide a check on whether reversed events have similar population characteristics as our candidates, which might indicate there is some instrumental issue worth considering that causes both. We have too few candidates here for such comparisons, but perform the search to get a sense of whether many reversed exocomet-like transits exist in the data.

We reversed the time axis for each light curve using $t_{\mathrm{rev}} = t_{\mathrm{final}} - t$, where $t_{\mathrm{final}}$ is the final timestamp for each light curve. The same vetting procedures described in Table \ref{table:candidates filtered} were then applied. This test was applied on the \texttt{eleanor-lite} and TESS-SPOC samples for sector 7.

Manual inspection of the surviving candidates showed that any reversed-asymmetry signals were primarily instrumental systematics or symmetric transit-like events, such as eclipsing binaries with unsubtracted background flux (that were confirmed with visual inspection of their TPFs). No convincing reverse-shaped candidates were identified. The main conclusion we can draw from this test is that reverse-shaped events are not significantly more common in our data than the forward events.

\section{Candidates} \label{sec: candidates}
After manual vetting, 6 candidates remain and are listed in Table \ref{tab:final candidates}, one of which is the known exocomet system $\beta$~Pic. The HR diagram in Fig \ref{fig:hr diagram} shows our candidates with the background colour showing the density of all stars that were searched. No reddening correction has been attempted for the full sample. We also obtained the following stellar properties for our candidates from Gaia DR3: stellar type, magnitude, parallax, angular separation from the Galactic plane, extinction ($A_g$ and $\mathrm{E}(B_\mathrm{p}-R_\mathrm{p})$), and effective temperature. The latter two quantities only exist where spectra were available, so we also fit stellar models to photometry as described below. The distances for the background stars were taken from the TIC, which is 96\% complete for our sample. The 4\% of the TICs in our sample did not have distances and therefore were omitted from the plot. This plot provides a test of whether exocomet candidates appear distinct from the parent search population; if candidates tend to be where the population density is also highest, the implication is that either all stars are equally likely to show exocomet-like behaviour, or that the events are not exocometary in origin, and instead are related to instrumental systematics and not related to stellar properties. There are two locations where detections have been recovered: the Main-Sequence (MS) branch, one candidate on the giant branch, and one candidate identified as a potential G-type supergiant. 

We fitted Spectral Energy Distributions (SEDs) for each of our candidates, and the derived parameters are shown in Table \ref{tab:final candidates}. The SEDs were modelled with a stellar atmosphere model with a modified blackbody to represent any dust component (see \citet{Yelverton2019,Yelverton2020} for details of the methods). We fit a single stellar model, so if the system is multiple the results represent either the more luminous primary, or both stars in the case of an equal-mass binary.
Aside from $\beta$~Pic, which has a known edge-on debris disc, none of these systems show any IR excess from our SED fits. From these fits, we can also estimate the radius, effective temperature, and reddening coefficient $A_v$ of the host stars, which are also listed in Table \ref{tab:final candidates}.

For a subset of these stars, Gaia BP/RP spectra were available and provided independent estimates of effective temperature and reddening. The results from the Gaia spectra and SED fits are in broad agreement with each other, and suggest that the reddening estimated by our SED fits are likely to be accurate, and would have flagged the two candidates without Gaia spectra as earlier types if they were significantly reddened. We briefly describe these results in the candidate descriptions, and the SED plots themselves are in Appendix \ref{appendix: SEDs}.

\begin{table*}
\centering
\resizebox{\textwidth}{!}{%
\begin{tabular}{|c|c|c|c|c|c|c|c|c|c|c|c|c|c|c|c|c|c|c|}
\hline
\multicolumn{7}{|c|}{\textbf{\textit{Candidate Properties}}} & \multicolumn{7}{c|}{\textbf{\textit{Stellar Properties}}} &  \multicolumn{4}{c|}{\textbf{\textit{SED Results}}} \\
\hline
Target & Sector & \makecell{Time\\(BTJD)} & \makecell{Depth \\ (\%)} & SNR & $\alpha$ & skew & Type & \makecell{$T$ \\(mag)} & \makecell{Distance \\ (pc)} & \makecell{E($B_{\rm p} - R_{\rm p}$) \\ (mag)} & \makecell{$A_{g}$ \\ (mag)} & b ($^{\circ}$) & \makecell{$T_{\rm eff}$ \\ (K)} & \makecell{$R_\star$ \\ ($R_\odot$)} & \makecell{$T_{\rm eff}$ \\ (K)} & \makecell{$A_v$ \\ (mag)} & Sp-type \\
\hline
TIC~270577175 ($\beta$~Pic) & 6 & 1486 & 0.15 & 5.94 & 1.02 & 1.44 & MS & 3.8 & 19.6 & 0 & 0 & -30.61 & 7938 & - & - & - & - \\
TIC~280832588 & 1 & 1344 & 0.27 & 5.23 & 1.05 & 3.20 & MS & 12.10 & 694 & 0.035 & 0.064 & -46.07 & 6516 & 1.4 & 6900 &  0.37 & F2V \\
TIC~73149665 & 12 & 1646 & 0.30 & 5.39 & 1.03 & 4.33 & MS & 11.69 & 417 & - & - & -01.38 & - & 1.54 & 5500 & 0.1 & $\sim$G7V \\
TIC~143152957 & 7 & 1499 & 0.55 & 6.70 & 1.04 & 3.14 & MS & 12.86 & 813 & 0.004 & 0.007 & -08.59 & 5503 & 1.64 & 5700 & 0.14 & $\sim$G6V \\
TIC~110969638 & 7 & 1513 & 0.39 & 6.24 & 1.05 & 6.43 & GIANT & 12.76 & 4348 & 1.230 & 2.269 & -03.08 & 5489 & 28 & 4200 & 1.41 & $\sim$G \\
TIC~229790952 & 15 & 1728 & 0.25 & 5.11 & 1.03 & 3.05 & GIANT & 10.99 & 1515 & - & - & 24.77 & - & 12.3 & 4700 & 0.17 & mid-K \\
\hline
\end{tabular}
}
\caption{Properties of the exocomet candidate transits and the stellar properties of the host systems. The stellar properties were taken from Gaia DR3, where distance was taken as 1/parallax as not all candidates had a calculated distance using Gaia DR3's \texttt{distance\_gspphot} parameter. The SED results are the derived stellar properties of the host star (see \citep{Yelverton2019,Yelverton2020} for details of the SED fitting methods). There is no SED value for $\beta$~Pic as their stellar properties are well-defined. The stellar properties here are well-constrained, and so the uncertainties are not included in the table. Where stated, the spectral types were estimated by comparing the derived effective temperatures with \citet{PeacutMamajek2013}.}
\label{tab:final candidates}


\end{table*}

\subsection{Candidates around Main-Sequence Stars}
\subsubsection{TIC~270577175 - $\beta$~Pic} \label{candidates: Beta Pic}
\begin{figure*}
\centerline{\includegraphics[scale=0.56]{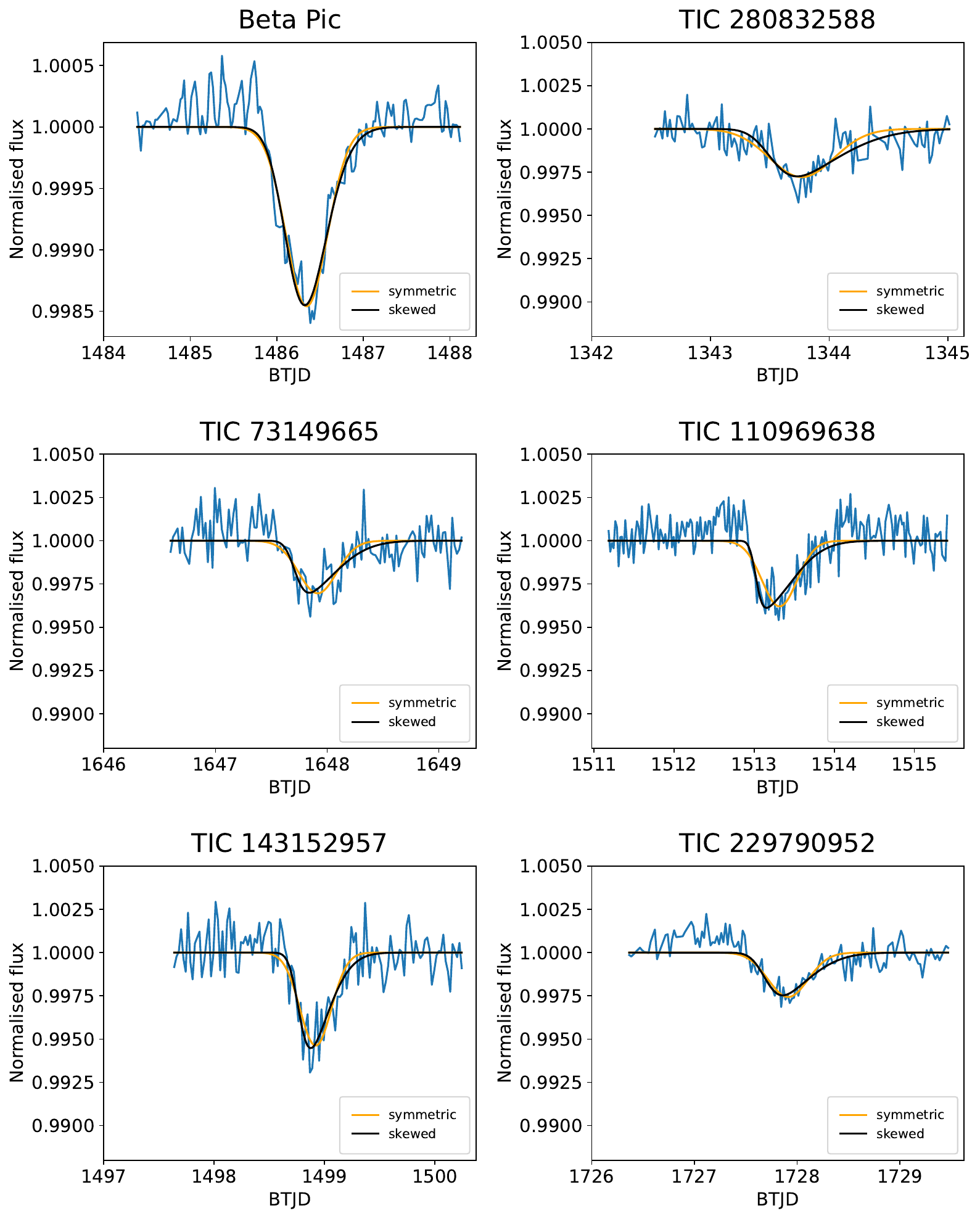}}
\caption{Light curves of the exocomet candidates centred on the transit time. The black line is the best-fit skewed Gaussian model, and the orange line is the symmetric Gaussian model. All panels apart from $\beta$~Pic share the same y-axis range to show their respective transit depths.}
\label{fig:candidates}
\end{figure*}

We successfully recovered the Sector 6 exocomet at $\beta$~Pic discovered in \cite{Zieba2019} at BTJD 1486, a reassuring sign that the detection methodology works as intended when applied to TESS data. $\beta$~Pic is an A6V star with a distance of 19.6 pc and a TESS magnitude of 3.8, making it a saturated target with TESS. The transit was only recovered in the TESS-SPOC light curves due to the saturation limitations of the \verb|eleanor-lite| light curve. We did not recover the exocomet events at BTJD 1459 (Sector 5) as it marginally did not pass our duration cut of 0.4d and the skewness cut of $<$ 8. In addition, the exocomet event at 1442 was also not recovered as it did not return an SNR greater than our SNR $>$ 5 threshold. These two shallower transits were originally detected with the aid of detailed modelling of the delta-Scuti pulsations with 2\,minute cadence data and subtraction of that model in \cite{Zieba2019}, and were not recovered by our method. There are detections of exocomets in $\beta$~Pic in the TESS extended mission \cite{Pavelnko2022,lecavalierdeseetangs2022}, but all of those detections are a factor of 10 shallower than the BTJD 1486 event, and would not be identified with our search method. The light curve of the recovered $\beta$~Pic transit can be seen in Fig \ref{fig:candidates}. 

\subsubsection{TIC~280832588}
TIC~280832588 is a $T = 12.1$\,mag star which showed a 0.27\% transit in Sector 1. The system was also observed in Sectors 2 and 3. One QLP light curve exists in the extended mission at Sector 28, and upon further inspection, a possible dimming event in S28 was concluded to be an unsubtracted stellar trend in the flux. This system is located at a distance of 694 pc and with an observed Gaia $B_{\rm p}-R_{\rm p}$ of 0.629 ($\sim$ F5V), the candidate sits just on the Zero Age Main Sequence (ZAMS, Fig \ref{fig:hr diagram}). 

The event appears to have happened a few days before a Sector 1 systematic issue across multiple light curves, and since our search method already accounts for events happening close to data gaps and there are no apparent issues with the light curve diagnostics, this is a robust event. There are three systems a few pixels away from the target on the TESS TPF. They are Gaia DR3 4695754063308036608, Gaia DR3 4695754269466468864, and Gaia DR3 4695754269466468608. No light curves are available for these targets, and we checked the PSF\_FLUX for the target to ensure there was no background contamination.

The candidate star was labelled in Gaia DR3 as an astrometric binary, with a Gaia Renormalised Unit Weight Error (RUWE) value of 4.97. The target is identified as having a non-linear proper motion compatible with an acceleration solution.
If the binary companion has a sufficiently long period, this could support the triggering of cometary orbits into the primary system with mechanisms such as the Kozai-Lidov mechanism \citep{Young2024}, but the inclination of the binary to trigger this process is unknown. The target was also processed by the TESS-SPOC pipeline, and we found that the target failed at the asymmetry score threshold of 1.02, and note that the light curve quality between the two pipelines could alter the result of the model fitting slightly. 

The SED results recover an effective temperature of $\sim$ 6900K, which is in broad agreement with the Gaia DR3 temperature, and suggests that the primary star in this system is an F2V star that is not significantly reddened (or if an equal mass binary then both stars are F2V and the radius of each is about Solar). While this transit is the only unique event in the data, and we see no signs of periodicity, without information on the binary companion, we acknowledge this as a tentative detection.

\subsubsection{TIC~73149665 (TYC 8705-361-1)}
TIC~73149665 is a $T = 11.7$\,mag star that showed a 0.3\% transit at BTJD 1647. The target is observed in Sector 12 and is located at a distance of 417 pc. There are several fainter stars in the nearby TESS pixels. After inspecting the pixel-by-pixel light curves, recovering the transit using the PSF flux, and ruling out periodicity by checking for detections in other sectors of TESS data in the extended mission, the event remains detectable. The star is only $1^{\circ}$ out of the Galactic plane and $35^{\circ}$ from the Galactic centre, so its location might suggest some reddening. The candidate has no GDR3 temperature, but its GDR2 temperature of the star is 5425K. However, as noted in \cite{Andrae2018}, this value is likely too low for significant extinction. The SED results suggest low reddening, with a derived $A_v = 0.1^{+0.1}_{-0.07}$, inferring a G-type main-sequence star. However, its location close to the Galactic plane motivates a better estimate of the level of reddening using spectra.


\subsubsection{TIC~143152957}
TIC~143152957 is a $T = 12.86$\,mag star that showed a 0.55\% transit in sector 7 at BTJD 1499. Gaia suggests that the target is a G-type with low reddening. The fitted SED results further suggest that there was not much reddening on this target, with $A_v = 0.14_{-0.11}^{+0.17}$, consistent with a G-type classification. However, the star is $8^{\circ}$ out of the Galactic plane and is located at a distance of $784\,$pc, so as with TIC~73149665, better stellar characterisation would be valuable. We inspected the target pixel file for possible contamination from neighbouring stars and while there are a few bright neighbouring stars a few pixels away, there were no associated dimming events in those systems, and it appears that the dimming event is occurring from the target. 

\subsection{Candidates around giant stars}
The remaining candidates, TIC~110969638 and TIC~2297970952, were detected around giant stars. There has only been one previous detection of an exocomet candidate around a giant star, which was through a spectroscopic survey \citep{RBW2024}, though the authors consider that the candidate more likely shows variation related to intrinsic stellar variability than an exocomet transit. It is potentially interesting to explore the nature of exocomets around more evolved stars. While the nature of exocomets around giant stars is still unclear, TIC 110969638 and TIC~229790952 passed all our cuts and manual vetting before analysing their stellar properties, so we treat them with an exocomet-like interpretation; that they are long-duration, low-amplitude transits with asymmetric transit profiles.


\subsubsection{TIC~110969638} 
TIC~110969638 is a Tmag = 12.76 star that showed a 0.39\% transit in Sector 7, at BTJD 1514. The TPF of this star is contaminated with a few background stars, containing three other stars with a $T_{\rm mag} > 13.7$, and TIC~110969585 ($T=11$) a few pixels away. Checking the pixel-by-pixel level light curves, the event appears strongest at the central pixels, and we can rule out the neighbouring stars as possible contamination. Verifying this with the PSF\_FLUX of the light curve to rule out possible contamination by the stars nearer to the target on the detector also recovers the transit with confidence. TIC~110969638 was observed twice again in the extended missions, in Sectors 34 and 61, with no signs of a repeated transit. 

The target has a Gaia $B_{\rm p} - R_{\rm p}$ of 2.101 before corrections. Applying the E($B_{\rm p} - R_{\rm p}$) and extinction corrections, the candidate's position on the HR diagram indicates that this is a G-type supergiant. The SED results also find a larger $A_v$ value of 1.41, also suggesting that the target is significantly reddened. G-type supergiants are rare, with less than 100 stars per pixel in this region of the HR diagram. The transit event in this light curve would either point to G-type giants commonly hosting exocomets, or that the event is better explained by other properties such as the intrinsic variability in this class of stars. Another possibility is that this transit originates from one of the contaminating stars from nearby pixels on the detector, but inspecting the light curves of the neighbouring stars does not show any transit signs from those systems, suggesting that it is more likely that the event is intrinsic to the star itself.

\subsubsection{TIC~229790952}
TIC~229790952 hosts a 0.25\% event detected at BTJD 1728 in Sector 15. The host system has an observed $B_{\rm p} - R_{\rm p}$ of 1.28, which makes it a K-type giant. There are no corrections for $B_{\rm p} - R_{\rm p}$ or extinction available for this target from Gaia DR3, making it difficult to suggest a ``corrected'' spectral-type. However, the low parallax suggests that this system is far away and likely reddened; however, the extent of the reddening may be less than TIC~110969638, as it is $24^{\circ}$ away from the Galactic plane. Indeed, the $A_v$ derived from the SED suggests a low reddening value of $0.17^{0.14}_{-0.09}$, but as with the other candidates, more robust values with spectral observations would be valuable to better constrain reddening and stellar parameters.

The transit event is interesting, as upon inspection of other TESS sectors, there appear to be a few other events. However, they have a lower SNR than the BTJD 1728 event and as such they did not pass through our cuts in the vetting process, and it is difficult to characterise their shape upon a visual inspection. A notable event is in Sector 17, where an apparent at BTJD 1768 does not appear asymmetric, but could be a shallower transit event as seen for $\beta$~Pic. The light curve for the Sector 17 event can be seen in Appendix \ref{Appendix: TIC 229790952}.

The system is well-observed, with 42 sectors of data available for this target as it is located in the TESS Continuous Viewing Zone (CVZ). There are no nearby stars on the detector that would cause any contamination, and with other false positive scenarios ruled out by checking their background fluxes and centroiding, the event points towards real astrophysical events around the star.



%

\begin{figure*}
\centerline{\includegraphics[scale=0.65]{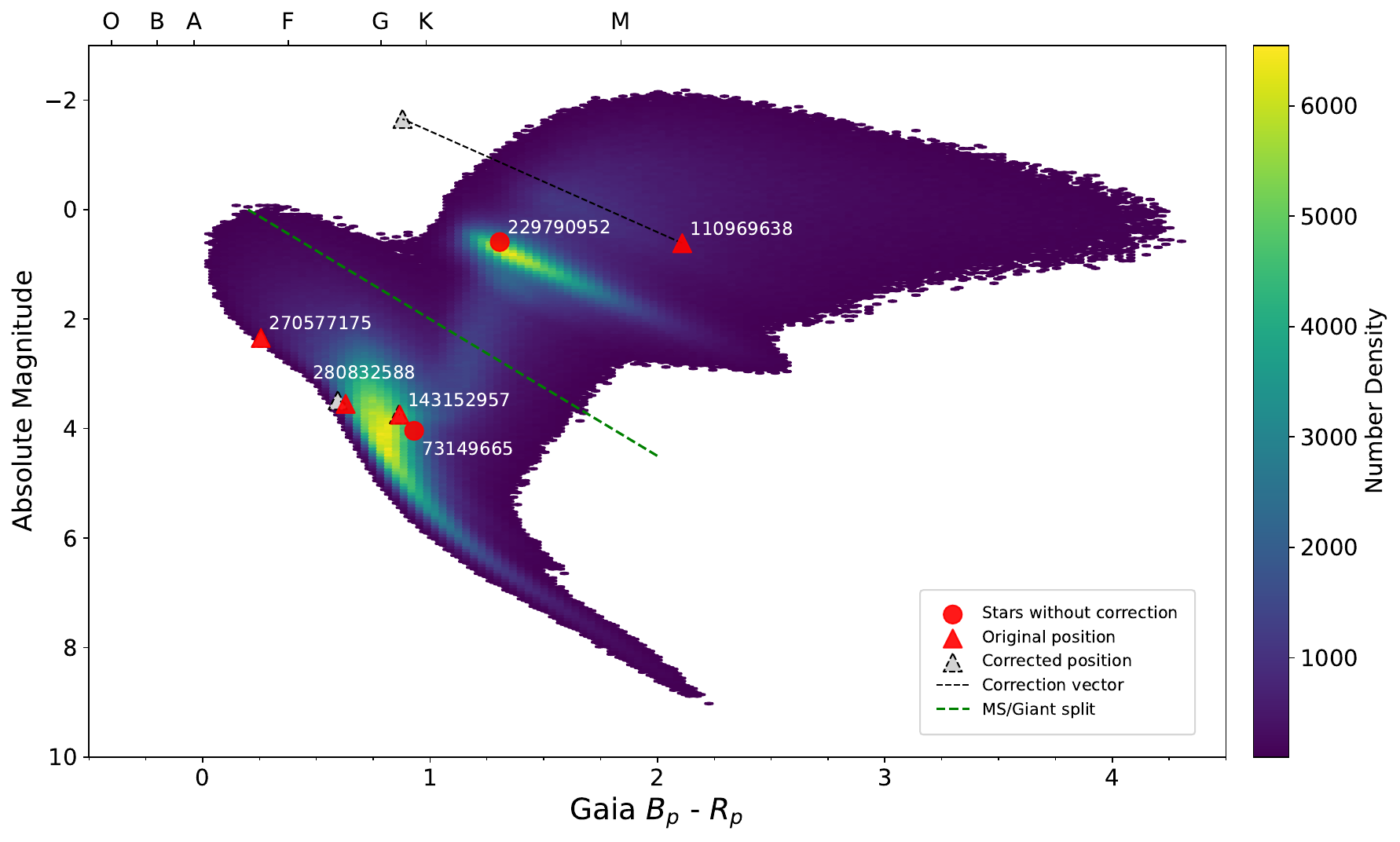}}
\caption{A Gaia HR diagram showing all stars with \texttt{eleanor-lite} light curves in our search sample, and our exocomet candidates overplotted on them. The background colour is the number of stars per bin from the search sample, where each bin contains a minimum of 100 stars. The background stars are not corrected for interstellar reddening. The red circles are the exocomet candidates for where no correction was required (i.e.: $\beta$~Pic) or no E($B_{\rm p} - R_{\rm p}$) and extinction values were available (i.e.: TIC~73149665 and TIC~229790952). For the candidates where E($B_{\rm p} - R_{\rm p}$) and extinction values were available, the red triangles show their original positions on the HR diagram, and the grey triangles show their positions after corrections for E($B_{\rm p} - R_{\rm p}$) and extinction.}
\label{fig:hr diagram}
\end{figure*}

\section{Discussion} \label{sec: discussion}

\subsection{Main-Sequence Candidates}
A fundamental question of exocomet science is whether exocomet detections are more likely around young, early-type stars compared to later spectral types and/or older stars, or if this apparent trend is due to observational biases that make detections easier around specific spectral types. Trends seen in spectroscopic observations suggest that exocomet detections are likely to be present around young, early-type stars. However, there is a bias in these results as observations have focused on systems with debris discs \citep{Montgomery2012,Iglesias2018,ExocometsRebollido}, which are easier to detect around younger, more luminous stars. Recent work in \cite{RBW2024} overcame this limitation in spectroscopy by conducting a survey using the entire HARPS archive, but still faces biases from the aims of the instrument: to search for exoplanets with radial velocities, where FGKM stars dominated the sample. On the other hand, the target selection of the large-scale photometric surveys so far (i.e. \textit{Kepler} and TESS) were not based on criteria that are directly related to debris discs or follow-up programs. 

Our candidates are located in different regions of the HR diagram from Fig \ref{fig:hr diagram}, where they are identified as main-sequence stars or giant stars. Two of our four main-sequence candidates are clearly A/F-type stars ($\beta$~Pic and TIC 280832588). The spectral type of the other two candidates is uncertain, but both are near the Galactic plane; TIC~73149665 has no observational reddening estimate, however the derived SED value for $A_v$ is only $0.1^{+0.1}_{-0.07}$, which may mean either very low extinction or exhibits the limitations of only photometric fitting for the SEDs. TIC~143152957 also has a perhaps surprisingly low reddening estimate from GDR3 and from our SED fits, where its proximity to the Galactic plane and a distance of 784 pc may mean that the reddening is underestimated. While better characterisation of all candidates is warranted, our SED and Gaia results were fairly consistent with each other where both were available, and so the G-type classifications are likely correct. We can conclude that exocomet candidates may also be found around later-type stars than previously indicated, though more detections will be needed to build a statistically significant sample.




$\beta$~Pic (TIC 270577175) is known to have an abundance of exocomet activity from spectroscopic observations \citep{ferlet1987,kiefer2014} and the presence of an edge-on debris disc. We note that we did not detect exocomets around any other systems with known debris discs (i.e.: the few hundreds of stars within several hundred pc with known debris discs \citep{Hughes2018}) as none of our candidates have infrared excesses. However, the deepest transit seen for $\beta$~Pic will be much more difficult to detect around the fainter stars that dominate our TESS sample, and that there are only of order hundreds of systems with known debris discs \citep{Chen2014,Cao2023}. Higher precision data for fainter stars, particularly younger stars with debris discs in young clusters/associations, could more strongly test whether $\beta$~Pic is anomalous among debris disc hosts, or if they are more common than seen from this work.

As TESS is an all-sky survey, its target list includes stars of various ages, though field stars should dominate the sample compared to young stars. One challenge in detecting exocomets around young stars is the handling of stellar activity. This is particularly challenging where rotational modulation is on a similar timescale to an exocomet transit of up to a few days long. In our large-scale search for exocomets, we removed slower variations with our median filter as discussed in Section \ref{subsec: choice of smoothing filter}, but while this filter method was successful in detrending slower variations, observing single transit events is tricky for faster-rotating variations as there is the risk of misidentifying stellar rotation as transit dips, or vice versa.





\subsection{Exocomets around Evolved Stars} \label{Exocomets around Evolved Stars}

TIC~110969638 and TIC~229790952 are the first photometric detections of potential exocomets around giant stars. We discuss briefly how planetesimals could survive the evolution of a star from the main sequence to its post-main-sequence phase.

Stars leaving the main sequence and ascending the giant branch become more luminous. They rise up the Red Giant Branch (RGB) before eventually moving to the horizontal branch. The star then rises on the Asymptotic Giant Branch (AGB), where it undergoes significant mass loss and ultimately ends up as a white dwarf system. During the giant phases, the sublimation radius can extend out to exo-Kuiper belt distances. The increasing stellar radiation from the more luminous star, coupled with the orbital expansion of bodies due to dynamical instabilities caused by mass loss, likely leads to the depletion of the cometary reservoir \citep{Stern1990}. The rate at which the comet reservoir is being depleted depends on the mass of the star during its main sequence stage \citep{veras2024}, but it is expected that the icy material will sublimate violently and stellar radiation would lead to rapid destruction of the dust. The comets on wide orbits then sublimate as they pass the extended sublimation radius into the inner regions of the system.

Debris discs lose mass over time, and observing events around giant stars suggests that a process is happening to increase the flux of comets to the star, or that a process occurs to make potential comets more visible. The dynamics of stars in the post-main-sequence stage can become more unstable depending on their stellar properties and system architecture, where two main processes can affect the orbital evolution and the survival of cometary bodies in this stage. The likely process is that in post-main-sequence systems, comets are perturbed onto eccentric orbits through gravitational interactions with larger bodies, where they would sublimate at much greater distances compared to the main-sequence stage due to the extended sublimation radius. The other consequence could be the Yarkovsky effect, which would propel planetesimals such as comets inwards or outwards on long timescales \citep{veras2024}. It is unclear from a single detection which of these consequences are in effect, but it could explain why comets can be detected around giant stars.

For our two detections, TIC~110969638 has a transit depth of 0.39\%, and TIC~229790952 shows a transit depth of 0.25\%, meaning that the events show a transit fraction similar to those of the MS systems. This could suggest that we are detecting much larger bodies compared to exocomets around main-sequence stars. With the variety of exocomet properties (i.e. orbital dynamics, dust production rate, composition etc), it is also possible that the coma is faced with more energy input from the more luminous giant star. This causes the nucleus to release more material and increase its size, affecting its observed transit depth. The conclusion for TIC~110969638 with the exocomet interpretation is unclear, and there could be an alternative explanation for the transit. The host is a G-type supergiant, which is rare in the parent sample seen in the Fig \ref{fig:hr diagram}. While we have ruled out the event as a false positive, verifying with Gaia DR3 that the host is a lone system, and inspecting the neighbouring stars for contamination, the rarity of G-type supergiants makes it challenging to draw conclusions about the frequency of exocomets around this stellar classification. If true, this would then suggest that exocomets should be common for G-type supergiants, but the event may be better explained by these events being an intrinsic variability in this stellar subtype. Further observations of this system or detections around other G-type supergiants could provide better context for a conclusive explanation. 

For TIC~229790952, the observed spectral type and the extinction values from our SED fits identify the candidate as potentially a mid-K, along with the majority of the giant sample. The case for this candidate appears to be a more straightforward explanation than TIC~110969638, where the candidate may have been detected as a result of interactions of the body approaching its host star. Across the 42 sectors of data that this target is observed, there are a few repeating transits as seen in Appendix \ref{Appendix: TIC 229790952}, though the photometric precision of TESS makes it challenging to characterise the shallower events. Future observations of this system would benefit from high-precision photometry to describe the true source of these events, and may be an interesting field star to explore the nature of exocomets around giants.

\subsection{Occurrence Rates} \label{occurrence rates}

Now that we have obtained a list of candidates and have performed a calculation of the detection efficiency of our pipeline with the injection-recovery tests in Section \ref{sec: injection recovery}, we can quantify occurrence rates of exocomet detection for the TESS primary mission. By drawing a diagonal line on the HR diagram seen in Fig \ref{fig:hr diagram} to split between the main sequence and giant candidates, our detection rate for main sequence candidates is $5.87\times10^{-7}$ star$^{-1}$\,sector$^{-1}$, and our detection rate for giants is $3.10\times10^{-7}$ star$^{-1}$\,sector$^{-1}$ (the units for the rate are per star per time, and we quote them as star$^{-1}$\,sector$^{-1}$ for TESS, or more generally star$^{-1}$\,year$^{-1}$). We ignore the spectral type and luminosity class in the occurrence rate calculations. In the following, we simply combine all candidates; with so few detections the uncertainties are very large, and our aim is to give an estimate of the detection rates rather than precise statistics. The occurrence rates can be calculated using the expression:


%
\begin{equation} \label{eq: occurrence rate}
    f_{\rm occ} = \frac{N_{\rm det}}{N_{\rm T}} \times \frac{1}{f_{\rm inj-rec}},
\end{equation}
where $f_{\rm occ}$ is the occurrence rate for a transit depth bin at a given magnitude, $N_{\rm det}$ is the number of exocomet candidates in that bin, $N_{T}$ the total number of light curves for stars observed at that magnitude, and $f_{\rm inj-rec}$ accounts for the detection efficiency through the our injection-recovery plots in Figure \ref{fig:injection_recovery} (b). The values of $f_{\rm inj-rec}$ decrease for shallower transit depths and fainter magnitudes, and the corrections therefore increase and can be large. Due to the small number of detections, we cannot compute true occurrence rates across the entire parameter space. However, we can estimate upper limits, for which we assume a detection of one candidate in each empty bin at a given depth and a given magnitude. We chose $N_{T}$ to be the total number of light curves instead of stars as our detection rate is expected to be proportional to the amount of time spent observing stars. Therefore, more observational data should yield more detections and more stringent upper limits. TESS continues to monitor the sky, and hence a search with subsequent TESS sectors could potentially find new candidates and populate more of this parameter space.







\begin{figure*}
\centerline{\includegraphics[scale=0.6]{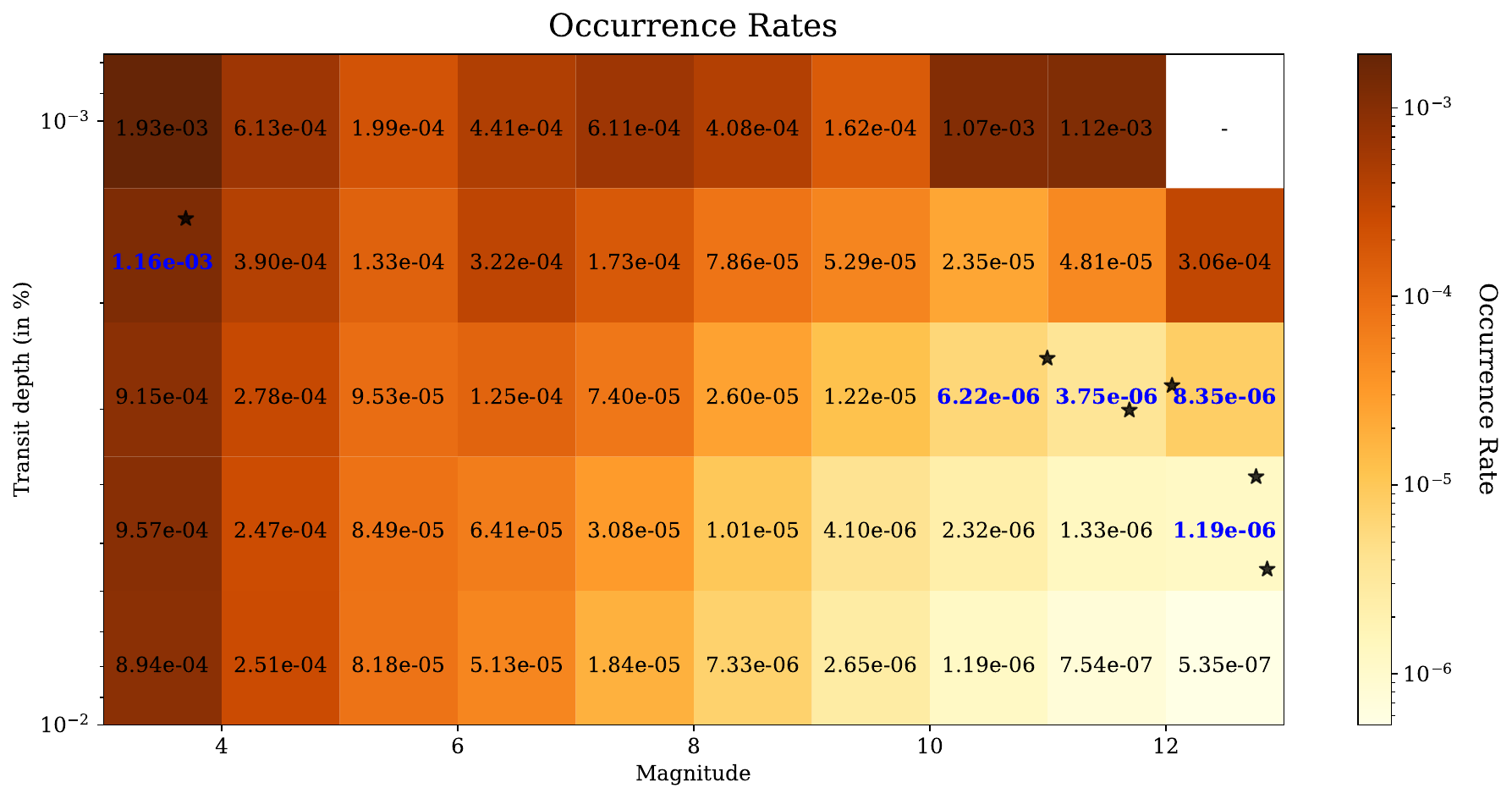}}
\caption{Estimations of exocomet occurrence rates in the TESS primary mission, where the detection rate values shown are for a given magnitude at given depth. The units for these occurrence rates are per star per TESS sector (star$^{-1}$\,sector$^{-1}$). The black crosses highlight the locations of our five exocomet candidates i.e.: the real occurrence rates of those bins, indicated with the blue text. The black text represents loose constraints on the occurrence rates based on 1 detection in each bin. The bins where there are no values are due to the detection efficiency in these bins being 0\%.}
\label{fig: occurrence rates}
\end{figure*}







Figure \ref{fig: occurrence rates} shows the occurrence rates at a given depth in a given magnitude range, where the detection efficiency used from Fig \ref{fig:injection_recovery} (b) is used for the calculation in Eq \ref{eq: occurrence rate}. Our search yielded 6 detections out of 15,477,322 light curves, giving an overall raw detection rate of $5.24 \times 10^{-6}$ per star per year, where the conversion from light curves per TESS sector to per year observations was done by multiplying the detection fraction by the ratio of days in a year to the $\sim$ 27-day sector observations in each light curve. This detection rate applies to transit depths down to 0.1\% before accounting for the detection efficiency. 

$\beta$~Pic is located on the leftmost region of the plot, with an occurrence rate of $1.16\times 10^{-3}$ star$^{-1}$\,sector$^{-1}$ (i.e. $1.53\times10^{-2}$ star$^{-1}$\,year$^{-1}$). The other candidates are all detected around stars towards the faint end of our magnitude range and have occurrence rates of order $1\times10^{-6}$ star$^{-1}$\,sector$^{-1}$ ($1.35\times10^{-5}$ star$^{-1}$\,year$^{-1}$). The injection recovery correction factor, which estimates the number of missed events due to the detection efficiency at these bins. From left to right in Fig \ref{fig: occurrence rates}, the correction factors are 1.86, 10.62, 18.69, 22.83 (for the occurrence rate of $1.19\times10^{-6}$ star$^{-1}$\,sector$^{-1}$), and 90 (for $8.35\times10^{-6}$ star$^{-1}$\,sector$^{-1}$), which emphasises the decreasing detection efficiency at fainter magnitudes. With only a few detections, the uncertainties on these rates are very large, and we conclude that exocomet transits at these depths are overall very rare.



The occurrence rate calculated for $\beta$~Pic is at least 1000 times higher than all the other candidates, marking it as an outlier in our candidate sample. The system is bright ($T=3.8$), has an edge-on debris disc, and is at a close distance of 19pc. Such a combination of features makes the $\beta$ Pic system the archetypal star for exocomet study. The clear lack of similar systems around fainter stars, where the upper limits in Fig \ref{occurrence rates} show that objects with 0.1\% transits with a $1\times10^{-3}$ star$^{-1}$\,sector$^{-1}$ occurrence rate are ruled out for $T=4$ to $T=13$ stars, further highlight that $\beta$~Pic is anomalous. In addition, HD 172555 is another system that has possibly yielded detections in both photometry and spectroscopy \citep{kiefer_2014_exocomets_hd172555,kiefer_cheops_hd172555_2023}. This system is thought to be in the young $\beta$~Pic moving group ($\sim$20 Myrs old) \citep{Zuckerman2001}, is at 29\,pc and also hosts a close to edge-on debris disc. The fact that these two systems have detections using both detection techniques may suggest that young moving groups appear favourable for exocomet detection. While HD 172555 was not observed in the TESS primary mission, its only light curve at the time of writing in Sector 66 did not show any signs of an exocomet transit. HD 172555 is expected to be observed again in Sectors 92 and 93, and future observations could provide us with the information needed to confirm any trends between exocomet detections and the young moving groups and disc orientations that they are associated with. We conclude that $\beta$~Pic's occurrence rate does not represent typical stellar systems, so we can estimate the occurrence rates excluding $\beta$~Pic. For the remaining detections, we have a total occurrence rate from summing the three bins of $1.95\times10^{-5}$ star$^{-1}$\,sector$^{-1}$, where $f_{inj-rec}$ is the average detection efficiency of the bins where there were detections. Multiplying by 365.25/27 results in an occurrence rate of $2.64\times10^{-4}$ star$^{-1}$\,year$^{-1}$ of exocomet transits that are 0.25\% or deeper. 





We can compare our results with the \textit{Kepler} search in \cite{Kennedy2019}. They returned a detection rate of approximately $7 \times 10^{-6}$ star$^{-1}$\,year$^{-1}$, where the occurrence rate values are a factor of 38 lower than our results. The main concern is that in both cases the rates are derived from a small number of detections, and so their uncertainties are similar to the rates themselves. There was no estimate of recovery efficiency for the \textit{Kepler} survey, which may have increased the rate, but the greater precision of \textit{Kepler} compared to TESS suggests that the effect is small. \textit{Kepler}'s photometric precision was an order of magnitude better than TESS \citep{Placek2016,Battley2021}, meaning that given an identical sample of stars, it should have yielded a higher rate than our results which was not the case. It is unclear why our TESS rate is significantly higher than the \textit{Kepler} rate, but it is clear that more detections are needed to better quantify the occurrence rates.

Aside from $\beta$~Pic, which appears to be anomalous, the number of detections is too small to cover much of the parameter space and quantify proper occurrence rate estimates. Therefore, the main conclusion for statistics is that photometrically transiting exocomets are extremely rare at this stage, but we can place initial constraints on the occurrence rates for a sample of magnitude-limited stars.


\section{Outlook} \label{sec: outlook}
\subsection{Photometric Precision}

A challenge in our search was the absolute photometric precision of the TESS light curves. While we carefully considered the requirements of our sample to match \textit{Kepler}'s star-years of observation in our injection recovery tests, the sensitivity coupled with our detection thresholds did not yield many transit detections compared to the transit depths observed in $\beta$~Pic, down to 0.1\%, for most other stars. The small number of exocomet detections is a limitation in understanding the phenomenon as only a couple of systems can be studied in depth. Although exocomets may be present in the TESS photometry for these systems, we would need significantly better photometric precision and/or longer observational baselines. 



Future missions such as PLATO \citep{Rauer2024} have the potential to detect and constrain many more exocomet-like transits. PLATO is expected to provide a 2-hour photometric precision of 7 ppm at $T = 4$ to $\sim$ 100 ppm at $T = 13$ (for targets observed with all 24 cameras) and an observation time of 2 years in its initial field, LOPS2 \citep{Rauer2024, Eschen2024, Nascimbeni2022}, leveraging both high-precision photometry with long observational baselines. Given that smaller objects are known to be more common in the Solar System than larger objects \citep{KenyonBromley2004,Morbidelli2020}, and that shallower transits are seen to be more common for $\beta$~Pic \citep{lecavalierdeseetangs2022}, our candidates were likely events towards the extreme end of the size distribution and missing a larger population of shallower transits. The increased photometric sensitivity of PLATO should be advantageous as, assuming all other things are equal, it increases the number of stars for which current transits at about 0.1\% could be detected, but also pushes farther down the size distribution to shallower transits for stars that have already been observed. Finding systems for which multiple comets are detected would allow us to make comparisons with the Solar System and $\beta$~Pic's planetesimal size distributions.


\subsection{Future work}

There are many open questions in exocomet science, where a fundamental question to address is the frequency of exocomet candidate detections around younger, early-type stars. Although a lack of stellar characterisation limits the strength of our conclusions, the main sequence candidates from our work and \cite{Kennedy2019,Rappaport2018} could suggest that this trend is a true phenomenon, which at a minimum provides a null hypothesis for future work. 

A consideration is whether $\beta$~Pic's observations have overestimated the detections for exocomets. The unique characteristics of $\beta$~Pic makes the system special in that exocomets were detected here before more common stellar systems. Most stars in the sky are not as active or bright as $\beta$~Pic, and our expectations of exocomet detections along with the photometric precision of TESS may still not be enough to probe most other young systems in close detail. Multiple young stars have shown exocometary activity with spectroscopy, so targeted searches around young systems with high-precision photometry could yield a better insight as to what is seen in $\beta$~Pic.

Another key consideration in exocomet searches is the efficiency and completeness of different search methods. While this work and \cite{Rappaport2018,Kennedy2019} have successfully identified candidates, it is worth examining to what extent such search methods are complete and have identified all potential signals. As noted earlier in Section \ref{subsec: choice of smoothing filter}, the detrending approach here is different from \cite{Kennedy2019} due to computational constraints and the specifics of TESS photometry. While necessary for processing our large dataset, our approach may have impacted our sensitivity to transit events as the surviving variability would have reduced their average SNR. Although lowering detection thresholds would identify more events, this also increases false positives given the TESS photometry's characteristics. So while our search methods both here and \cite{Kennedy2019} have proven to work as a useful recovery tool for known exocomet transits and yields a number of new candidates, the potential to fully explore exocomet transits could also be through alternative search techniques. Furthermore, while our focus on only the most significant signals in each light curve which was effective at eliminating most systematics and noise, this meant that we potentially missed multiple smaller transit events such as the two transits detected in Sector 5 of $\beta$~Pic \citep{Zieba2019}, and therefore our occurrence rates being a lower bound to the true rate. 



The TESS mission is still ongoing and presents an opportunity to explore new data, with approximately 2.6 million more star-years of observations based on the QLP target list from sectors 27-82 up to $T = 13$, and hence potentially new exocomet detections. This also presents the opportunity to revisit previously observed systems, and explore new detection methods such as machine learning (ML), where there has been considerable success in exoplanet transit detection e.g: \citet{ShallueVanderburg2018,Ansdell2018,osborn2020}. Given the complexity of exocomet transits and the large volume of unexplored data, ML methods could be promising. Recent work in \citealt{Dobrycheva2024-exocomets-ML} developed a method using Random Forests (RF) based on statistical features of light curves, using the photometric precision as an indicator for injecting exocomet signals. In addition, we are currently exploring Convolutional Neural Networks (CNNs) as a new technique, which may be promising as the ML models are being trained on the exocomet shapes themselves. These alternative approaches may complement existing methods to provide a more complete census of the exocomet phenomenon. 

The primary objective for future photometric surveys continues to be determining the most effective ways to detect and characterise exocomet detections with high-precision photometry. Novel techniques to characterise transit shapes and the upcoming PLATO mission could provide this opportunity to explore stars with the precision needed to yield new candidates around stars fainter than $\beta$~Pic.

\section{Conclusions}
\label{sec: conclusions}

We presented a search for exocomet transits in the TESS primary mission to determine their occurrence rates as a function of spectral type and stellar age. We reported five new candidates, along with the recovery of the exocomet event detected around $\beta$~Pic from \cite{Zieba2019}. The search algorithm is based on the work from \cite{Kennedy2019}, where we searched for single transit events in a light curve, and used cuts to narrow down our sample based on the known/expected characteristics of exocomet transits (i.e. duration, depth, skewness, etc). We then visually inspected candidates that passed all our thresholds, yielding the 6 candidates from a list of 272. 

We find that TIC~280832588 is an F-type main-sequence star, consistent with the spectral-type range as the previous exocomet candidate detections. However, the other two main sequence candidates, TIC~73149665 and TIC~143152957, both show evidence of being G-type stars based on their SED fits and Gaia data, where available. This would mean that these candidates represent the first detections around later-type stars, and if confirmed, these are the first Sun-like stars with exocomet detections. However, spectroscopic observations would be valuable for definitive stellar classification.



The remaining candidates, TIC~110969638 and TIC~229790952, were found around giant stars. TIC~110969638 was found around a G-type supergiant, a rare subclass in our parent sample, and TIC~229790952 was observed around a mid-K giant before corrections. While debris discs can theoretically exist on the giant branch, little is known about them. In this case, giant stars could show exocomet transits due to the effects of mass loss on dynamics and/or the changes in stellar luminosity. 


We computed occurrence rate estimates for the detections, where we get a total occurrence rate of $2.64 \times 10^{-4}$ star$^{-1}$\,year$^{-1}$. This is higher than \textit{Kepler}'s detection rate of $6.7 \times 10^{-6}$ star$^{-1}$\,year$^{-1}$ \citep{Kennedy2019}. While our work has yielded new candidates, this is still in the realm of small number statistics to precisely quantify occurrence rates for photometric detections, and that the difference in occurrence rate is likely due to the the small number of detections and the difference in photometric precision between the two missions. We therefore conclude that more detections are needed to better constrain occurrence rates of photometric transits.


We may have missed potential candidates in some cases. One instance would be for active young stars, where the stellar variability is a dominant feature of the light curves and raised as a potential transit event, then discarded as a false positive. The limitations of automated techniques require the testing of different techniques and the assessment of whether alternative methods can yield exocomet candidates more efficiently than the search methods tested so far. Furthermore, the upcoming PLATO mission should yield the high-precision data required for new exocomet detections and provide valuable statistics.



\section*{Acknowledgements}

The authors acknowledge valuable discussions with Paul Strøm, Matthew Battley, and Edward Bryant. AN is supported by the University of Warwick and the Royal Society. GMK was supported by the Royal Society as a Royal Society University Research Fellow. This work is also supported by the National Aeronautics and Space Administration (NASA) under grant No. 80NSSC21K0398 issued through the NNH20ZDA001N Astrophysics Exoplanets Research Program (XRP). The contributions by SG have been supported by STFC through consolidated grants ST/P000495/1, ST/T000406/1 and ST/X001121/1. Computing facilities were provided by the Scientific Computing Research Technology Platform at the University of Warwick.

\section*{Data Availability}

The TESS data used in this work are publicly available on the Mikulski Archive for Space Telescopes (MAST) portal \url{https://mast.stsci.edu/portal/Mashup/Clients/Mast/Portal.html}. The TESS-EB catalogue can be shared upon reasonable request. The code for this project is available on \url{https://github.com/azibn/automated_exocomet_hunt}.



\bibliographystyle{mnras}
\bibliography{references} 




\appendix

\section{Test-statistic of the candidates}
Similar to Fig \ref{fig: search-method}, the figures in this section show how the search method identified our other candidates.

\begin{figure*}
\centering{\includegraphics[scale=0.55]{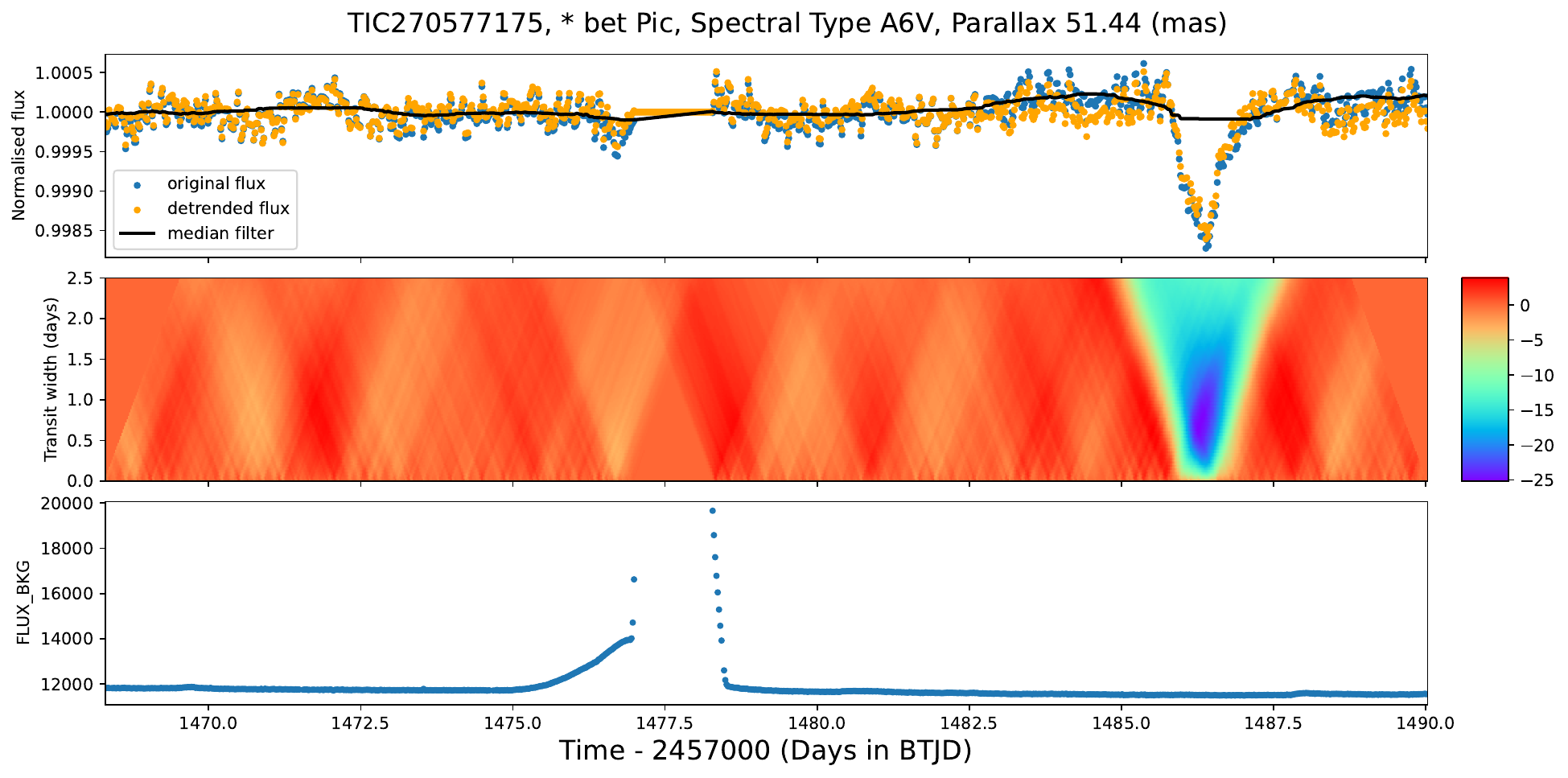}}
\caption{The light curve of $\beta$~Pic (top panel), its corresponding test-statistic (middle panel), and the background flux activity of the light curve (lower panel).}
\label{appendixfig:270577175}
\end{figure*}

\begin{figure*}
\centering{\includegraphics[scale=0.57]{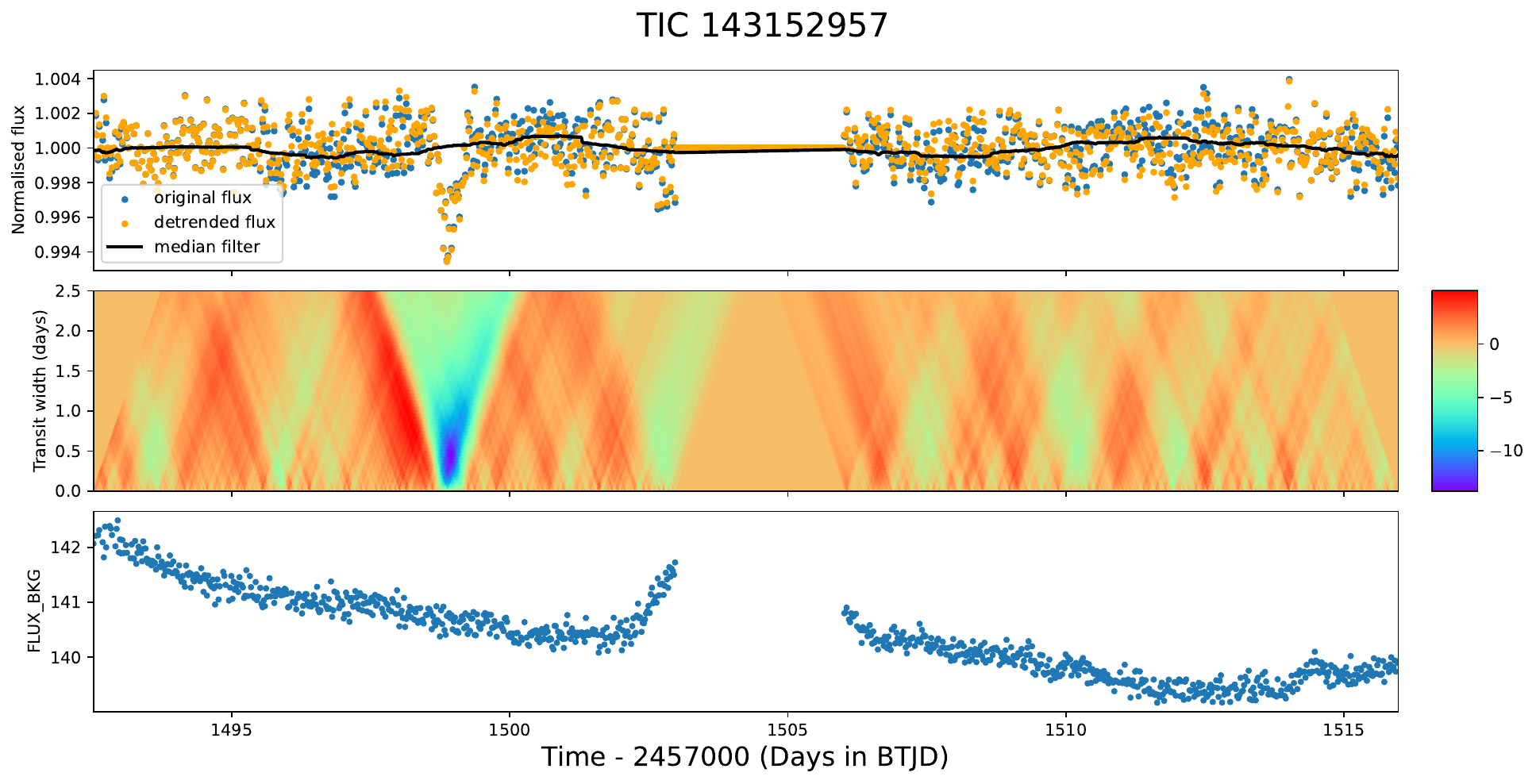}}
\caption{The light curve of TIC~143152957 (top panel), its corresponding test-statistic (middle panel), and the background flux activity of the light curve (lower panel).}
\label{appendixfig:143152957}
\end{figure*}

\begin{figure*}
\centering{\includegraphics[scale=0.57]{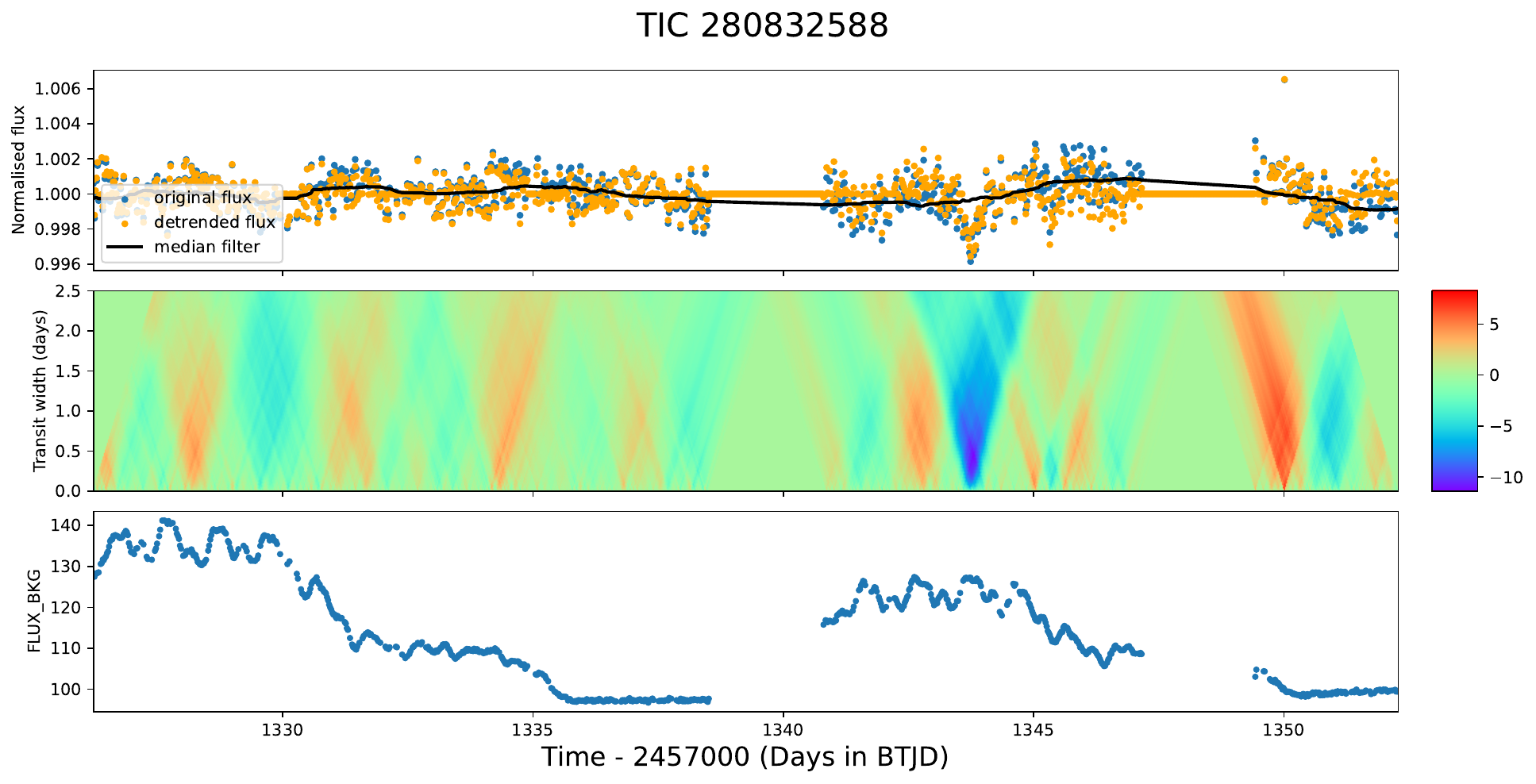}}
\caption{The light curve of TIC~280832588 (top panel), its corresponding test-statistic (middle panel), and the background flux activity of the light curve (lower panel). The background flux is active for this target, but this is not a concern as it appears to be consistent with a longer-term trend, and not related to the transit event.}
\label{appendixfig:280832588}
\end{figure*}

\begin{figure*}
\centering{\includegraphics[scale=0.57]{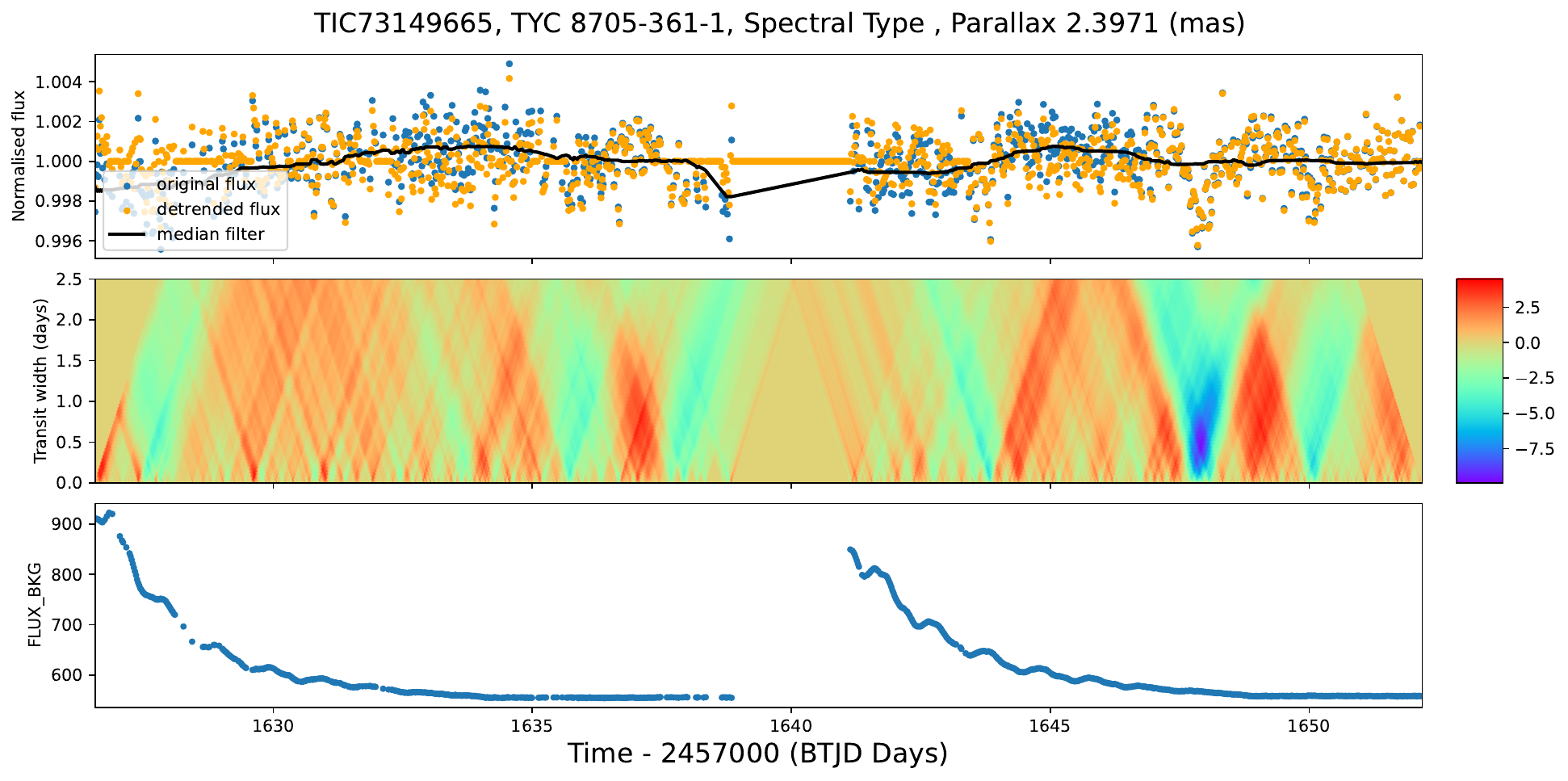}}
\caption{The light curve of TIC~73149665 (top panel), its corresponding test-statistic (middle panel), and the background flux activity of the light curve (lower panel).}
\label{appendixfig:73149665}
\end{figure*}

\begin{figure*}
\centering{\includegraphics[scale=0.57]{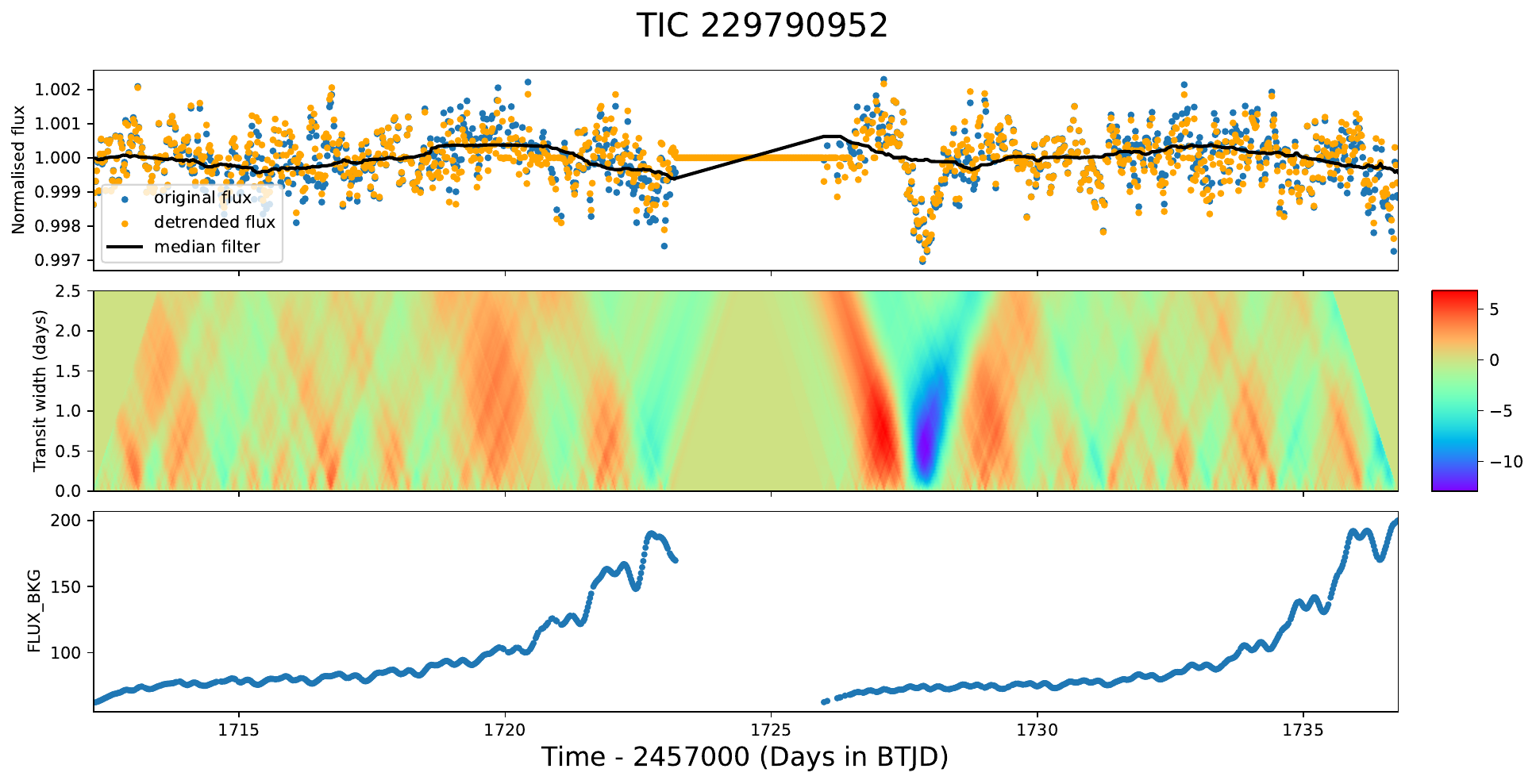}}
\caption{The light curve of TIC~229790952 (top panel), its corresponding test-statistic (middle panel), and the background flux activity of the light curve (lower panel).}
\label{appendixfig:73149665}
\end{figure*}

\section{Ruling out $\alpha$~Pic}\label{Appendix: ruling out alpha pic}
\begin{figure*}
\centering{\includegraphics[scale=0.48]{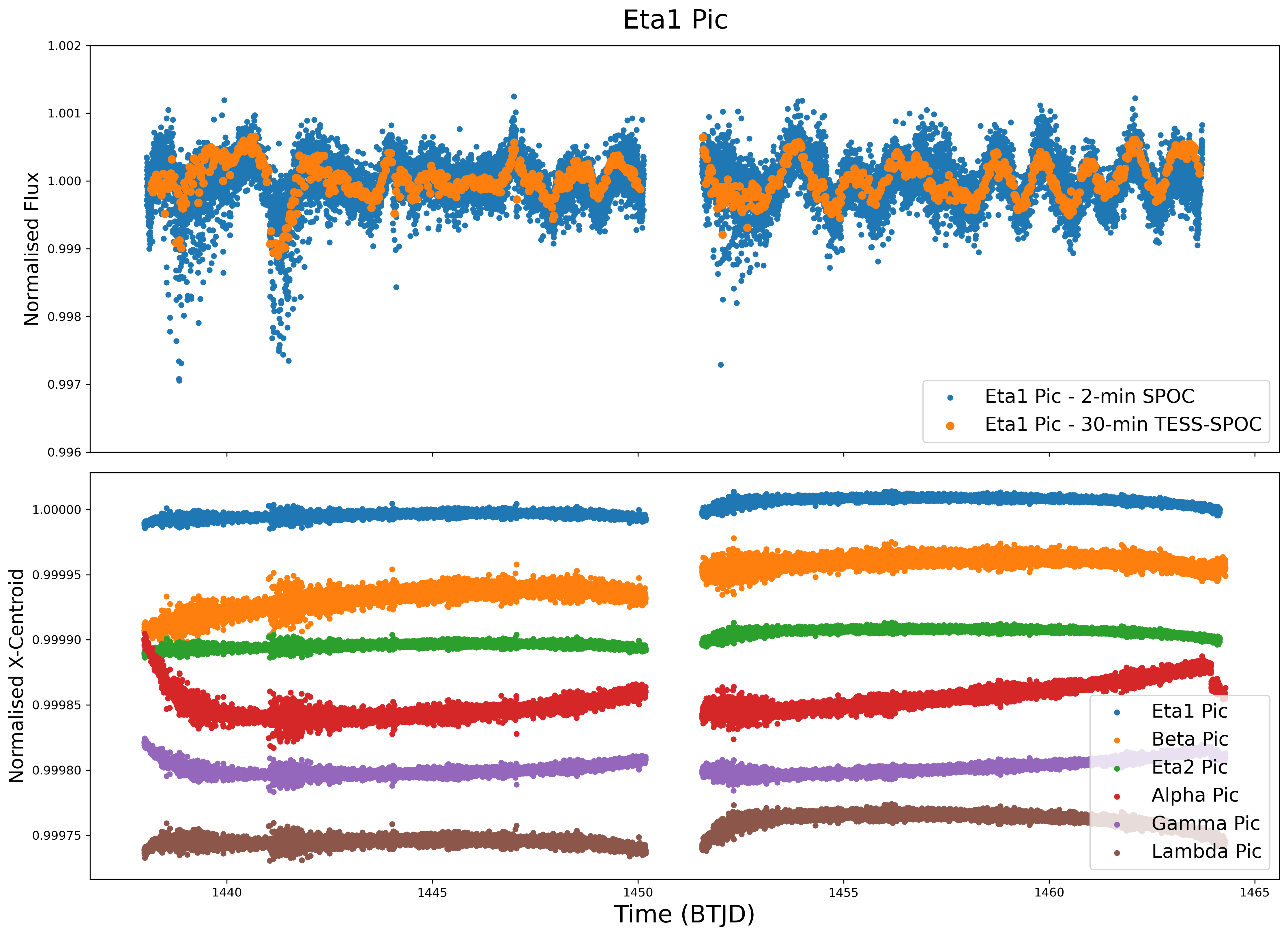}}
\caption{The centroiding columns in the selected light curves from the Pictor constellation to compare with $\eta^1$~Pic. There is an evident jitter at BTJD 1441 in the centroiding for all stars which corresponds to the event seen in the flux columns of the light curve.}
\label{fig: centroiding jitters}
\end{figure*}

Fig \ref{fig: centroiding jitters} shows the scatter in the centroiding of 2-minute cadence light curves for the selected targets in the Pictor constellation, where the activity in the centroiding was seen at the same time as the event in the $\alpha$~Pic light curve.

\section{Other light curves of TIC~229790952} \label{Appendix: TIC 229790952}
This appendix shows additional light curves of TIC~229790952 where a visual inspection identified a potential transit that did not pass all the cuts of our search pipeline.

\begin{figure*}
\centering{\includegraphics[scale=0.56]{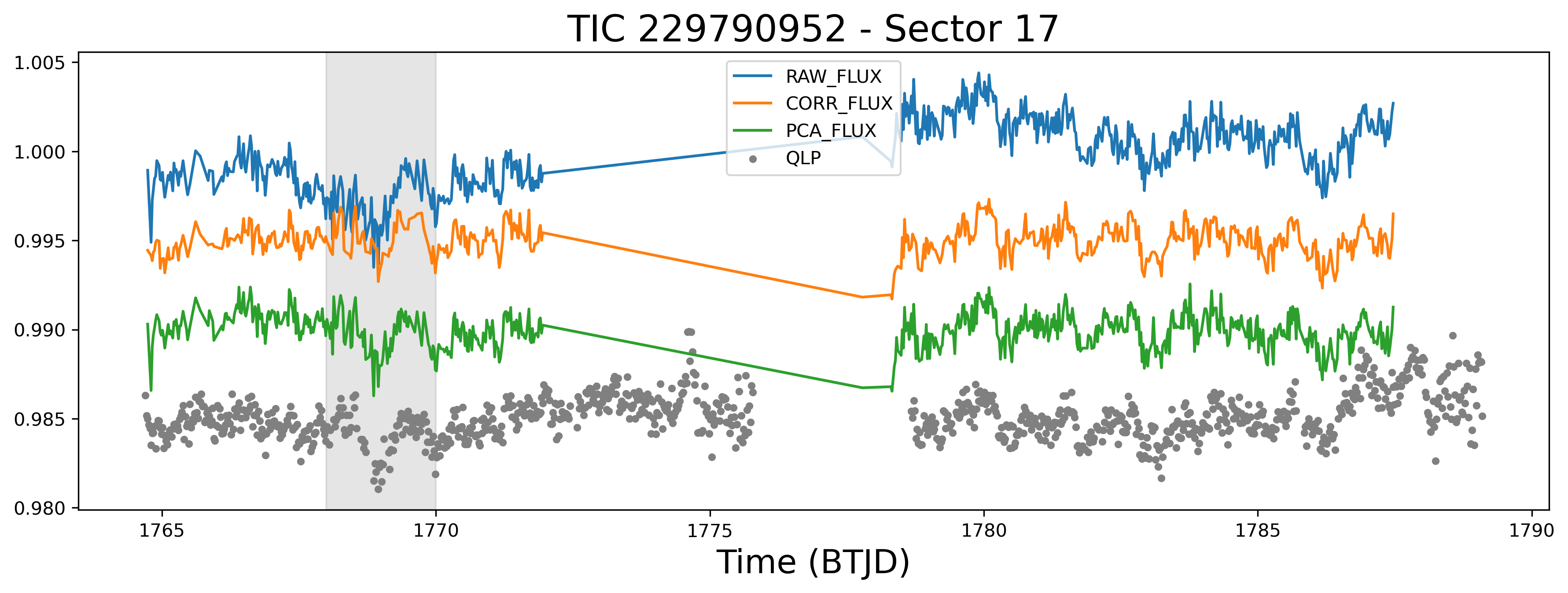}}
\caption{The light curve of TIC 229790952 in Sector 17, where there appears to be a dimming event at BTJD 1769 across the \texttt{eleanor-lite} data products and the QLP light curve. The QLP light curve shows the signal more prominently than the \texttt{eleanor-lite} light curves, and the validity of the event is unclear.}
\label{fig: tic229790952_s17}
\end{figure*}

\section{Candidate SEDs} \label{appendix: SEDs}
We show here the Spectral Energy Distributions (SEDs) for each of our new candidates. The SEDs were fitted using \texttt{sdb}, as described by \citet{Yelverton2019}.

\begin{figure}
\centering{\includegraphics[scale=0.56]{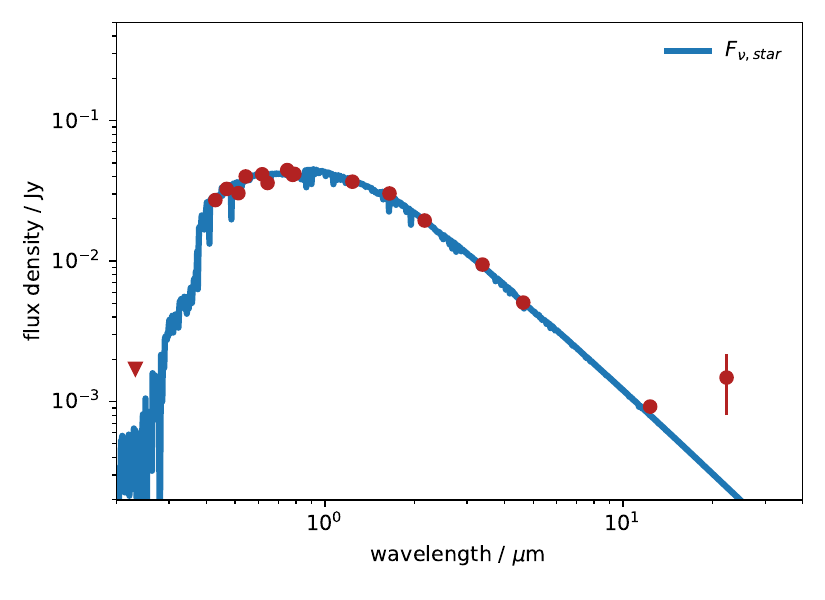}}
\caption{The SED for TIC 280832588.}
\label{fig: sed_280832588}
\end{figure}

\begin{figure}
\centering{\includegraphics[scale=0.56]{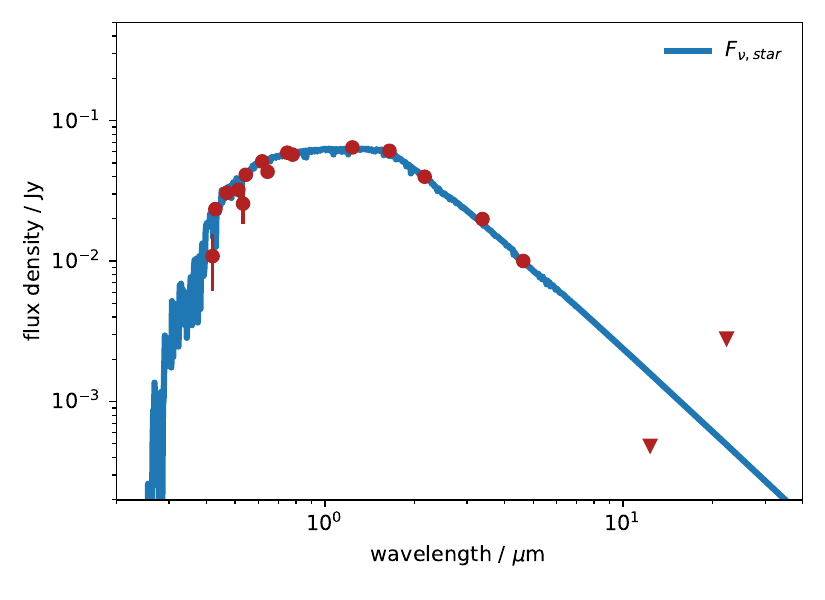}}
\caption{The SED for TIC 73149665.}
\label{fig: sed_73149665}
\end{figure}

\begin{figure}
\centering{\includegraphics[scale=0.56]{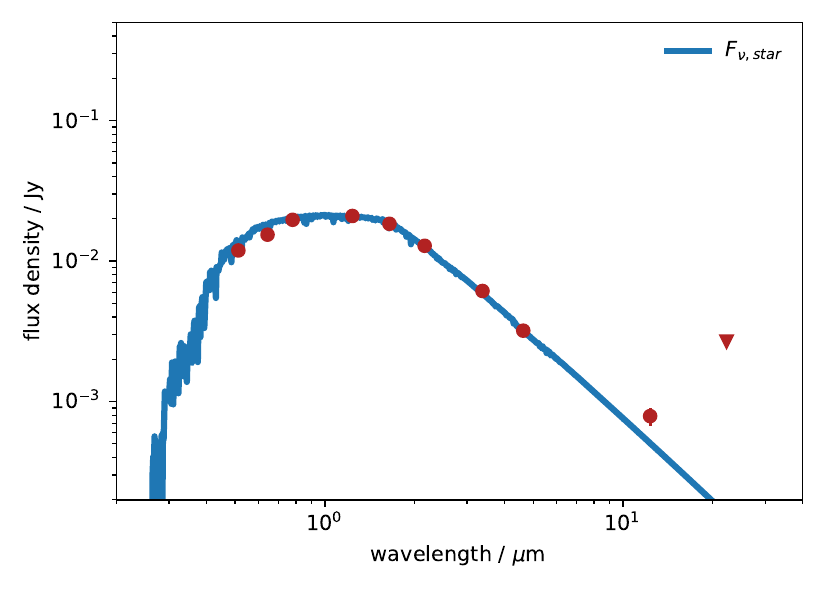}}
\caption{The SED for TIC 143152957.}
\label{fig: sed_143152957}
\end{figure}

\begin{figure}
\centering{\includegraphics[scale=0.56]{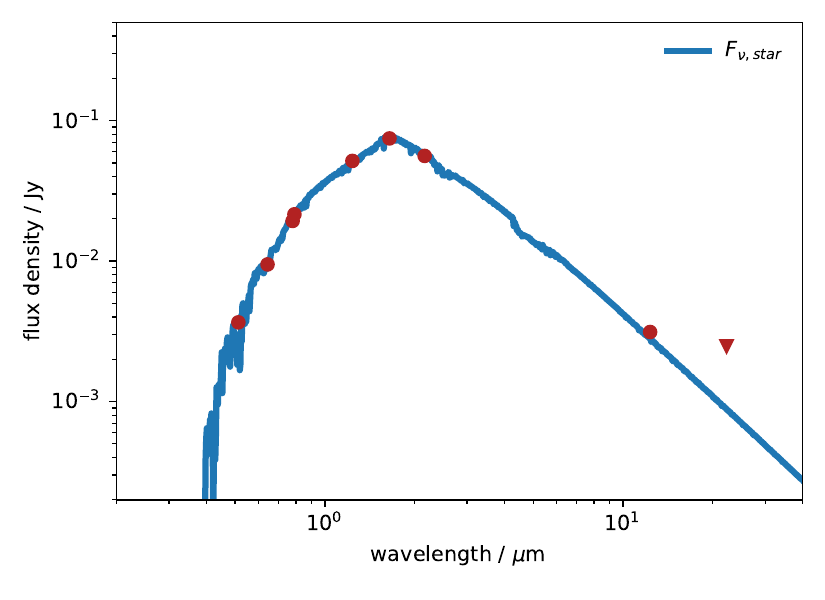}}
\caption{The SED for TIC 110969638.}
\label{fig: sed_110969638}
\end{figure}

\begin{figure}
\centering{\includegraphics[scale=0.56]{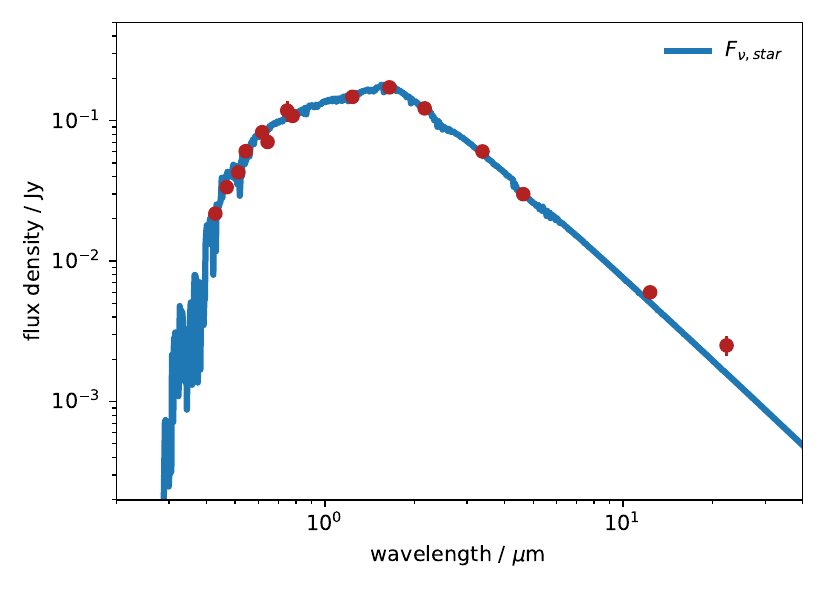}}
\caption{The SED for TIC 229790952.}
\label{fig: sed_229790952}
\end{figure}

\pagebreak
\bsp	
\label{lastpage}
\end{document}